\documentclass{article}
\usepackage{listings}
\usepackage{hyperref}
\usepackage[ruled,vlined,linesnumbered,algo2e]{algorithm2e}
\usepackage{subcaption}
\usepackage{caption}
\usepackage{enumitem}
\usepackage{microtype}
\usepackage{graphicx}
\usepackage{xcolor}
\usepackage{booktabs} 
\usepackage{multirow}

\usepackage[preprint]{neurips_2025}
\usepackage{neurips_2025}
\usepackage{amsmath}
\usepackage{amssymb}
\usepackage{mathtools}
\usepackage{amsthm}
\theoremstyle{plain}

\theoremstyle{definition}

\theoremstyle{remark}

\usepackage{titlesec}
\titlespacing*{\section}{0pt}{0.1\baselineskip}{0.1\baselineskip}
\titlespacing*{\subsection}{0pt}{0.1\baselineskip}{0.1\baselineskip}
\titlespacing*{\subsubsection}{0pt}{0.1\baselineskip}{0.1\baselineskip}

\usepackage[utf8]{inputenc} 
\usepackage[T1]{fontenc}    
\usepackage{hyperref}       
\usepackage{url}            
\usepackage{booktabs}       
\usepackage{amsfonts}       
\usepackage{nicefrac}       
\usepackage{microtype}      
\usepackage{xcolor}         
\usepackage{wrapfig}

\title{Spatial-RAG: Spatial Retrieval Augmented Generation for Real-World Geospatial Reasoning Questions}

\author{%
  Dazhou Yu\thanks{Equal contribution.} \\
  Emory University \\
  \And
  Riyang Bao\footnotemark[1] \\
  Emory University \\
  \And
  Ruiyu Ning \\
  Emory University \\
  \And
  Jinghong Peng \\
  \And
  Gengchen Mai \\
  University of Texas at Austin \\
  \And
  Liang Zhao\thanks{Corresponding author.} \\
  Emory University \\
}

\begin{document}

\maketitle

\begin{abstract}

Answering real-world geospatial questions—such as finding restaurants along a travel route or amenities near a landmark—requires reasoning over both geographic relationships and semantic user intent. However, Existing large language models (LLMs) lack spatial computing capabilities and access to up-to-date, ubiquitous real-world geospatial data, while traditional geospatial systems fall short in interpreting natural language. To bridge this gap, we introduce Spatial-RAG, a Retrieval-Augmented Generation (RAG) framework designed for geospatial question answering. Spatial-RAG integrates structured spatial databases with LLMs via a hybrid spatial retriever that combines sparse spatial filtering and dense semantic matching. It formulates the answering process as a multi-objective optimization over spatial and semantic relevance, identifying Pareto-optimal candidates and dynamically selecting the best response based on user intent. Experiments across multiple tourism and map-based QA datasets show that Spatial-RAG significantly improves accuracy, precision, and ranking performance over strong baselines.

\end{abstract}

\section{Introduction}
\begin{wrapfigure}{r}{0.5\linewidth}
  \vspace{-10pt}  
  \centering
  \includegraphics[width=1\linewidth]{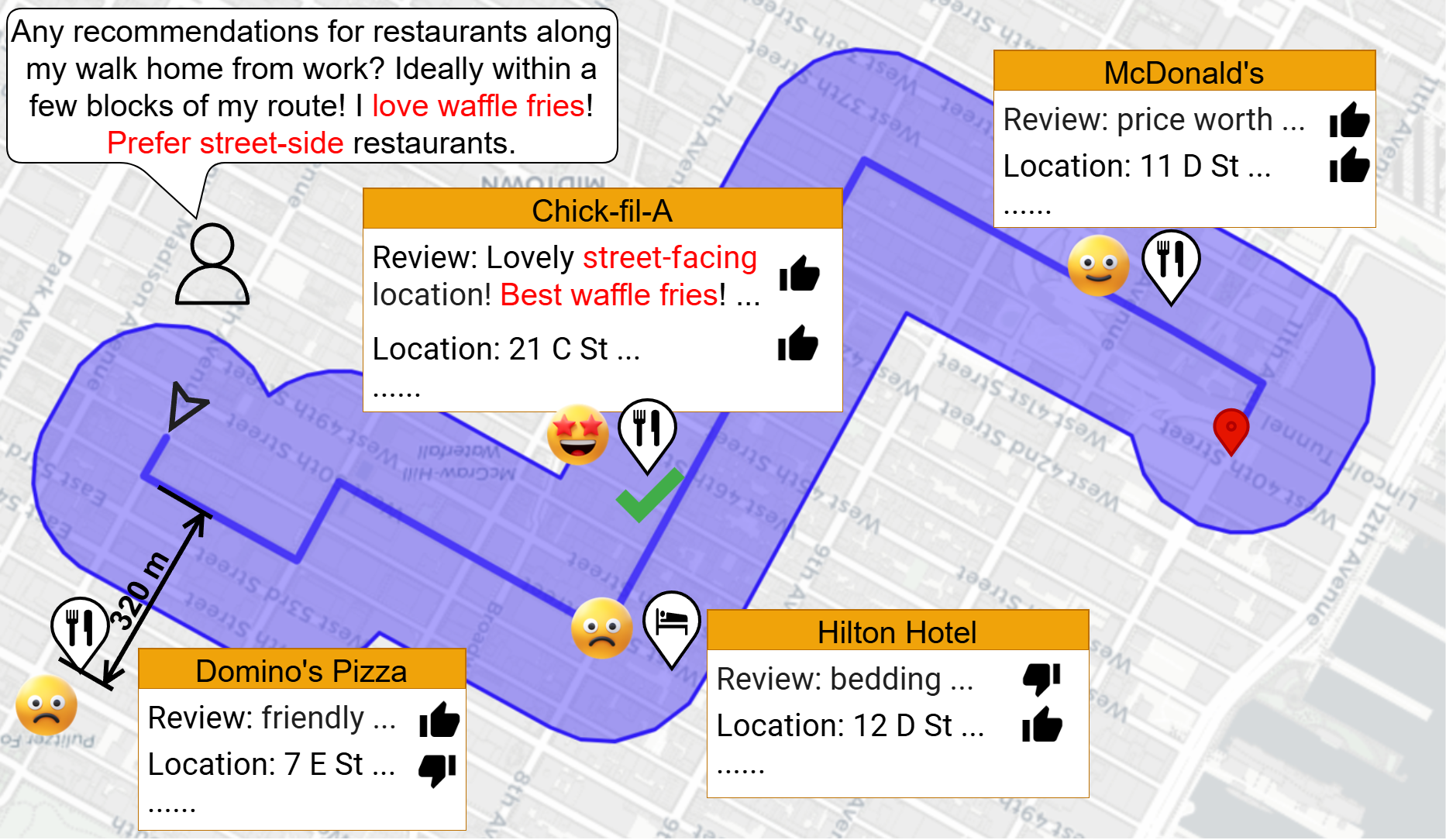}
  \caption{A real-world spatial reasoning question with nearby spatial objects. Areas that satisfy the spatial constraint are highlighted in purple.}
  \vspace{-0.3cm}
  \label{fig:intro}
\end{wrapfigure}

Spatial reasoning questions are those that require spatial computing to resolve relationships between objects, positions, or movements in space. Extensive research has been conducted on abstract spatial reasoning tasks such as mental rotation, block manipulation, and robot navigation, which rely on simplified, small-scale, and often purely geometric representations, typically addressed using techniques from computer vision and robotics. In contrast, geospatial reasoning involves interpreting large-scale, real-world geographic data where spatial information is deeply entangled with rich semantics \cite{chen2014parameterized,mai2021geographic,kefalidis2024question}. For example, urban routing decisions depend not only on road geometries but also on attributes such as traffic regulations, land use, and temporal constraints. Travel plan recommendations should not only consider minimizing the travel distance, but also maximize the quality of the attractions according to their descriptions and reviews \cite{xietravelplanner}.

Geospatial reasoning has a longstanding role in AI research, yet classical methods—such as spatial databases and GIS query systems—lack the ability to effectively interpret users’ natural language questions \cite{mai2021geographic}. On the other hand, large language models (LLMs) exhibit strong linguistic competence but struggle with spatial computing and geospatial grounding \cite{mai2024opportunities}. Recent efforts to bridge this gap have focused on prompt engineering \cite{manvigeollm,gurnee2024language}, but these approaches heavily rely on LLMs’ internal knowledge, which remains limited in generalization and spatial reasoning capabilities, significantly suffering from geographic bias \cite{faisal2023geographic,manvi2024large,wu2024torchspatial}, and being susceptible to obsolescence as knowledge evolves. Some work has explored fine-tuning LLMs on spatial tasks \cite{ji2023evaluating,manvigeollm,zhang2024bb}, but the resulting models are often tailored to narrow applications, constrained datasets, or specific geographic domains. Therefore, there remains a critical need for a general-purpose geospatial reasoning framework that \textbf{synergizes semantic understanding and spatial computation} while ensuring access to \textbf{real-world, vast, fast-changing, and complex geospatial data}.

To fill this gap, this paper aims to augment LLMs with capabilities of spatial reasoning and accessibility to real-world geospatial data. For example, as illustrated in Figure \ref{fig:intro}, answering the question requires LLMs to elicit and formulate the user's textual request into the problem of ``finding points near the polyline" and solve it based on a geospatial map (database) with semantic information (e.g., customer reviews and location profiles). Then, it also requires inferring user intent to select the spatially and semantically preferred candidates. 
Thus, the system must seamlessly integrate structured spatial retrieval with unstructured text-based reasoning, ensuring both spatial accuracy and contextual understanding. Specifically, we extend Retrieval-Augmented Generation (RAG) into geospatial information retrieval and reasoning, bridging the gap between structured spatial databases and unstructured textual reasoning. 
RAG has demonstrated its effectiveness in knowledge-intensive tasks, such as question answering (QA) \cite{siriwardhana2023improving}, by retrieving domain-specific documents to enhance LLM responses. 
However, existing RAG systems primarily focus on retrieving and generating textual content and lack the spatial intelligence required for spatial reasoning tasks, especially tasks that involve understanding and computing complex spatial relationships among geometries, including points, polylines, and polygons.  


In this paper, we introduce \textit{Spatial Retrieval-Augmented Generation (Spatial-RAG)}, a new framework that unifies text-guided spatial retrieval with spatially aware text generation under multi-objective optimization scenario. Specifically, to identify spatially relevant candidate answers, we propose a novel spatial hybrid retrieval module synergizing spatial sparse and dense retrievers. 
To rank the candidates and generate the final answers, we propose to fuel the generator with retrieved results on the Pareto frontier based on a spatial and semantic joint ranking strategy.
Our contributions are summarized as follows:
\vspace{-0.3cm}
\begin{itemize}[leftmargin=*,wide]
\setlength\itemsep{-0.2em}
\item \textbf{A generic spatial RAG framework}: We introduce spatial-RAG, the first framework that extends RAG to geospatial question answering, to tackle a broad spectrum of spatial reasoning tasks, such as geographic recommendation, spatially constrained search, and contextual route planning. Our approach seamlessly integrates spatial databases, LLMs, and retrieval-based augmentation, enabling effective handling of complex spatial reasoning questions directly within the familiar operational paradigm of LLMs.
\item \textbf{Sparse-dense spatial hybrid retriever}: We propose a hybrid retrieval mechanism that combines spatial sparse retrieval (e.g., SQL-based structured queries) with spatial dense retrieval (e.g., LLM-powered semantic matching). This dual approach ensures that retrieved results align both spatially and semantically with the user’s query, synergizing spatial computing and geographical text understanding.
\item \textbf{Multi-objective guided spatial and semantic text generator:} To handle both spatial constraints and semantic intents in the spatial question-answering task, we introduce a multi-objective optimization framework that dynamically balances trade-offs between spatial and semantic relevancy to the user's query. This ensures that the generated responses are both geospatially accurate and linguistically coherent.
\item \textbf{Real-world evaluation:} 
We evaluated our method on multiple real-world datasets consisting of user-generated QA pairs about various spatial entities. The experiments demonstrate the model’s ability to handle spatial reasoning questions grounded in real-world scenarios.
\end{itemize}

Through these innovations, Spatial-RAG significantly enhances the spatial reasoning capabilities of LLMs, bridging the gap between structured spatial databases and natural language QA.

\section{Related work}
\subsection{Retrieval augmented generation}
Retrieval-Augmented Generation (RAG) is a hybrid approach that integrates retrieval systems and generative models to enhance factual accuracy and contextual relevance in natural language generation \cite{fan2024survey}. Unlike conventional language models that rely solely on parametric memory, RAG dynamically retrieves relevant external knowledge before generating a response.
In RAG \cite{lewis2020retrieval},  a retrieval module fetches relevant passages from a large-scale knowledge corpus (e.g., Wikipedia), which are then fused with the question context to generate a more informed response. This technique has proven particularly effective in open-domain question answering (QA), fact verification, and context-aware text generation.
RAG systems have expanded beyond text and document retrieval to incorporate 
a wide variety of data types \cite{he2024g} — tables, graphs, charts, and diagrams. 
While RAG has been widely explored, its application in spatial reasoning question answering remains an unexplored research area. Existing studies have primarily focused on knowledge-grounded dialogues \cite{yu2024llms} but often struggle with integrating spatial computation into the question-answering process effectively.
\subsection{Geospatial question and answering}
Spatial questions in domain-specific applications can generally be categorized into two distinct types:
\textbf{1) Textual knowledge-based spatial questions} These are spatial questions that can be answered by traditional QA methods without the need for spatial computation and reasoning \cite{lietard2021language}.
For example, the question \textit{"What is the population of Los Angeles city?"} falls under this category. Despite their spatial context, these questions are essentially text-based and, hence, can be effectively addressed using traditional Retrieval-Augmented Generation (RAG) methods \cite{christmann2024rag}.
\textbf{2) Spatial reasoning questions} This category encapsulates spatial questions that demand a model's capability to comprehend and reason with spatial data and spatial relationships. A common example is a model being presented with textual information describing the spatial relationships among multiple objects \cite{li2024advancing}. An example question could be, \textit{"What is the position of object A relative to object B?"}, where objects \textit{A} and \textit{B} are locations or entities specified on the map. Resolving such queries requires a profound understanding of spatial concepts and robust reasoning skills, which largely depend on the model’s training to handle spatial data.
Several studies  \citep{mai2024opportunities,roberts2023gpt4geo} have investigated the capacity of LLMs to understand spatial concepts, yet these models often struggle with accurate reasoning even after fine-tuning.   Other research \cite{li2023geolm} has attempted to enhance this ability by converting geolocation coordinates into addresses to enrich the semantic context.   However, these improvements tend to be marginal and are mostly limited to straightforward reasoning tasks like describing positions. Moreover, many existing methods rely on predefined sets of actions tailored to specific tasks.

\section{Problem formulation}
In this study, our primary focus is 
\textbf{Geospatial Reasoning Questions}. We formulate the problem as follows:
Given a query $q$, the system aims to generate an answer $y^*$, which maximizes the joint spatial and semantic scores while satisfying the spatial constraints in the query $q$. For example, in Figure \ref{fig:intro}, the desired answer is a restaurant that satisfies a spatial constraint—it must be within walking distance of the route. Additionally, it should ideally be located along the street (spatial score) and preferably offer waffle fries (semantic score). This problem can be formulated as the following multi-objective optimization problem:
\begin{equation}
\begin{aligned}
y^* &= \arg \max_{y} \ \lambda_s^T f_s(q, y) + \lambda_k^T f_k(q, y)  \\
\text{s.t.} \quad & y \in C_s(q), \quad y \in C_k(q), \;
  \lambda_s \geq 0,   \lambda_k \geq 0, &
 \mathbf{1}^T \lambda_s + \mathbf{1}^T \lambda_k = 1,
\end{aligned}
\label{eq:problem_improved}
\end{equation}
where $f_s\in \mathbb{R}^{d_s}$ is the spatial relevance score vector, $f_k\in \mathbb{R}^{d_k}$ is the semantic relevance score vector, $C_s$ is the spatial candidate set that satisfies the spatial constraints of the question, $C_k$ is the semantic candidate set that satisfies the semantic constraints of the question, $\lambda_s, \lambda_k$ are the spatial weights and semantic weights, respectively,  $y^*$ is the optimal answer, $\mathbf{1}^T\lambda_s + \mathbf{1}^T \lambda_k = 1$ ensures a normalized trade-off. 

Note that this problem can have multiple valid solutions depending on the trade-off parameters $\lambda _s$ and $\lambda_k$, forming a \emph{Pareto frontier}. Existing approaches are unable to solve this problem effectively, as it demands a synergistic capability in both geospatial and semantic reasoning. Specifically: 1) it requires accurately determining whether a candidate $y$ satisfies the spatial constraints expressed in the query $q$. 2) it involves ranking candidates based on both spatial and semantic relevance, which are interrelated yet provide complementary signals; and 3) it must ensure that no high-quality answers are overlooked under different trade-offs between spatial and semantic aspects. Current methods, which typically excel in either geospatial reasoning (e.g., spatial databases) or semantic reasoning (e.g., large language models and their variants), cannot address all of these requirements without substantial effort to integrate both capabilities seamlessly. 

\section{Spatial-RAG for geospatial reasoning questions}

\begin{figure*}[htb]
  \centering
  \includegraphics[width=\textwidth]{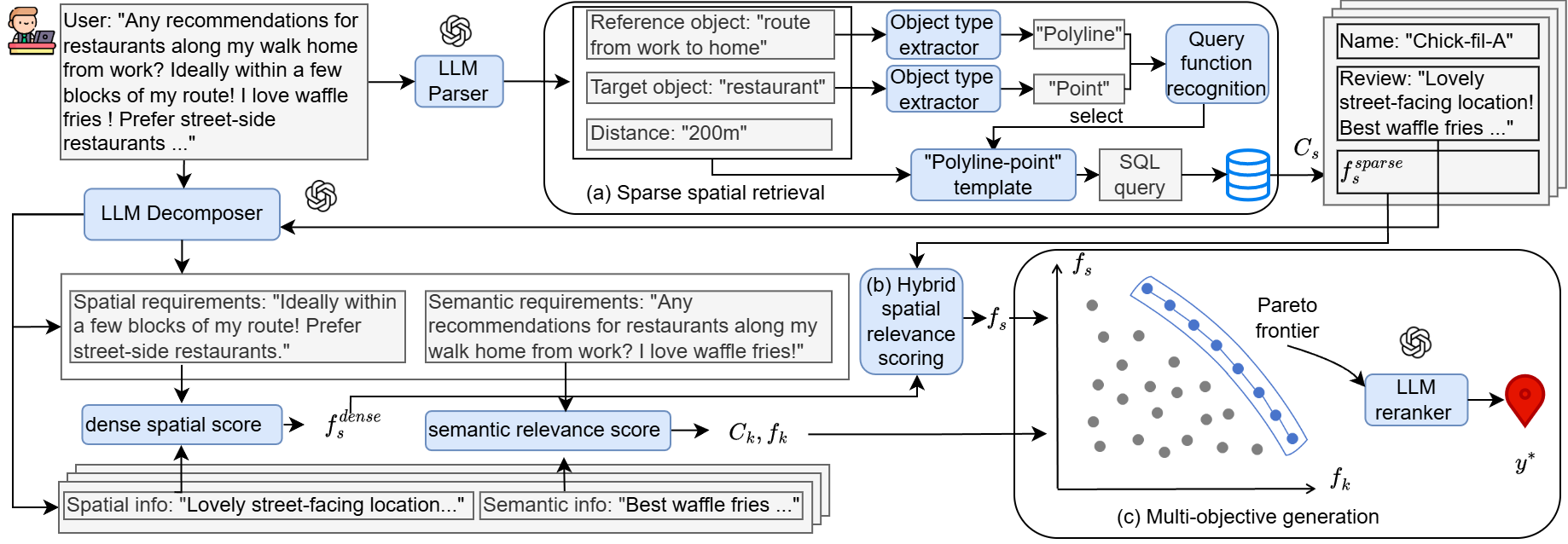}
  \caption{Illustration of the proposed Spatial-RAG framework.}
  \vspace{-0.5cm}
  \label{fig:arch}
\end{figure*}

\subsection{Overview}
Our Spatial-RAG is illustrated in Figure \ref{fig:arch}, which consists of three key stages:
First, to construct the spatial candidate set $C_s$, the system must precisely define spatial constraints and then retrieve spatial objects that satisfy them. As depicted in Figure \ref{fig:arch} (a) Sparse spatial retrieval, we achieve this by parsing the input natural language questions into a spatial SQL query, which will be executed on the spatial database to efficiently retrieve relevant spatial objects from the database. This process is detailed in Section \ref{sec:sql}.
Second, to effectively compute spatial relevance $f_s(q,y)$ while integrating textual information, we propose a hybrid spatial retrieval scheme. This method combines sparse spatial relevance scores from the database with dense semantic similarity scores from text embeddings. This enables the system to rank retrieved spatial objects based on their spatial relevance to the input question, as detailed in Section \ref{sec:ranking}.  
Third, given both spatial and semantic scores, we formulate a multi-objective optimization problem to balance these factors. The system computes the Pareto frontier of candidate answers, and the LLM dynamically trades off among these solutions to generate an optimal response. This step is covered in Section \ref{sec:generation}.


\subsection{Sparse spatial retrieval}\label{sec:sql}
The answer to a spatial reasoning question must meet specific spatial constraints. The spatial candidate set \( C_s(q) \) consists of all possible answers \( y \) that satisfy a set of spatial constraints \( \mathcal{C}_s(q) \): 
\begin{equation}
C_s(q) = \{ y \mid {c}_s(y, q) \leq 0, \forall {c}_s \in \mathcal{C}_s(q) \},
\end{equation}
where 
${c}_s(y,q)$  represents a constraint function that encodes a spatial condition (e.g., topological, directional, or distance-based constraints),
$\mathcal{C}_s(q)$  is the set of all spatial constraints associated with the question $q$.
For example, if the spatial constraint requires \( y \) to be within a distance \( \epsilon \) from a reference location \( l_q \), then a possible constraint function is:
\begin{equation}
c_s(y, q) = \text{d}(y, l_q) - \epsilon \leq 0.
\end{equation}
This formulation ensures that only spatially valid answers are included in \( C_s(q) \).

Addressing spatial constraints requires executing a well-defined spatial SQL query within a spatial database. This process involves identifying the appropriate query function, the reference spatial objects, the target spatial objects, and any necessary numerical parameters. Formally, a spatial SQL query can be expressed as:
\begin{equation}
Q_s = \mathcal{F}_s(G_r, G_t, \epsilon)
\label{eq:sql}
\end{equation}
where
\( \mathcal{F}_s \) is the spatial query function that determines the relationship between objects.
\( G_r \) represents the set of reference objects extracted from the question.
\( G_t \) represents the set of target objects that are potential answers.
\( \epsilon \) is the set of numerical parameters and spatial relationships governing the spatial constraint (e.g., distance threshold, topological relations).

Given the diversity and complex nature of these constraints, LLMs often struggle to directly construct a complete and executable spatial query from user input. To bridge this gap, we parse the spatial query incrementally, allowing LLMs to systematically populate the required components.
Our approach follows three key steps:
1) \textbf{Geometry recognition:} Identify and extract the reference spatial objects \( G_r \) and candidate target spatial objects \( G_t \) from the user’s input and extract their spatial footprints -- geometries.
2) \textbf{Query function selection:} Determine the appropriate spatial function \( \mathcal{F}_s \) based on the intended spatial relationship (e.g., containment, proximity).
3) \textbf{Parameter estimation:} Assign numerical constraints \( \epsilon \) to ensure precise spatial filtering (e.g., buffer radius).

By formalizing this structured process, we enhance the LLM’s ability to generate accurate and executable spatial SQL queries. This, in turn, improves the system’s capability to handle complex spatial reasoning questions effectively. 

\subsubsection{Geometry recognition}

In spatial reasoning tasks, accurately identifying spatial objects and extracting their spatial footprints (i.e., geometries) are essential for parsing questions to spatial queries. Spatial footprints of spatial objects, denoted as \( g\in\mathcal{G} \), can generally be categorized into three fundamental types: points, polylines, and polygons. Formally, we define these categories as follows:
\vspace{-0.3cm}
\begin{itemize}[leftmargin=*,wide]
\setlength\itemsep{-0.2em}
    \item \textbf{Point:}  
    \(    \mathcal{G}_\text{point} = \{ g \mid g \in \mathbb{R}^2, \dim(g) = 0 \} \)
    This category includes single points and multipoints, representing locations with negligible area. Examples include stop signs, points of interest, and a user's current location. In spatial databases, these entities are typically represented as the 'Point' geometry type.

    \item \textbf{Polyline:}  
    \(    \mathcal{G}_\text{line} = \{ g \mid g \subseteq \mathbb{R}^2, \dim(g) = 1 \} \)
    Polylines, including multipolylines, represent linear one-dimensional objects with negligible width. Common examples include streets, streams, bus routes, and power lines. In spatial databases, these geometries are abstracted as the 'LineString' type.

    \item \textbf{Polygon:}  
    \(    \mathcal{G}_\text{polygon} = \{ g \mid g \subseteq \mathbb{R}^2, \dim(g) = 2 \}   \)
    Polygons, including multipolygons, represent two-dimensional objects that define enclosed areas. These geometries are essential for depicting regions such as census areas, parcels, counties, neighborhoods, and zoning areas.
\end{itemize}
\vspace{-0.3cm}
The complexity of a spatial query depends on the types of spatial footprints of objects involved. For simpler queries, such as "finding the nearest bus stop from a given location", only point geometries are required, and the spatial candidate set is
\begin{equation}
    C_s =\{ g \vert g \in \mathcal{G}_\text{point}, d(g, g_\text{point}) < \epsilon \}
\end{equation}
where \( g_\text{point} \subseteq \mathcal{G}_\text{point} \) represents a point object (e.g., given location), $\epsilon$ is the distance threshold. 
For more complex queries, such as "I will walk from home to the university campus along 7th Street and Jones Street; please recommend a café where I can buy breakfast on my walk.", multiple geometry types must be considered, and the spatial candidate set is
\begin{equation}
    C_s =\{ g \vert g \in \mathcal{G}_\text{point}, g \in B(g_\text{polyline},\epsilon)\cup g_\text{polygon}\}
\end{equation}
where \( g_\text{polyline} \subseteq \mathcal{G}_\text{polygon} \) represents a polyline object (e.g., a route), \( g_\text{polygon} \subseteq \mathcal{G}_\text{polygon} \) represents a polygonal region (e.g., a university campus), $B$ is a buffer around the polyline, $\epsilon$ is the buffer size. 
 
By structuring spatial queries in this way, we ensure precise geometric representation, facilitating robust spatial reasoning and query execution.

\subsubsection{Query function recognition and parameter estimation}
After recognizing the geometries involved in a spatial query, the subsequent step is to determine the appropriate spatial query functions $\mathcal{F}_s$ required to handle various geometrical interactions. Despite the differing interactions among geometries, these can be uniformly addressed using distance functions $\text{d}(g_r, g_t)$, which calculate the shortest distance between two geometrical entities $g_r, g_t \in \mathcal{G}$.

Formally, given sets of reference geometries $G_r \subseteq \mathcal{G}$ and target geometries $G_t \subseteq \mathcal{G}$ , the spatial candidate set $C_s$ can be defined as:
\begin{equation}
\begin{cases} 
\{g_t \in G_t \mid \exists g_r \in G_r, \text{d}(g_r, g_t) \leq \epsilon \} , &\text{if }\text{d}(g_r, g_t) > 0, \\
\{g_t \in G_t \mid \exists g_r \in G_r, g_r \cap g_t \neq \emptyset \}, &\text{if } \text{d}(g_r, g_t) = 0.
\end{cases}
\label{eq:distance_query}
\end{equation}
Parameters such as search radius or buffer distance $\epsilon$ are autonomously determined by the LLM, typically grounded in contextual understanding (e.g., estimated walking distance or area of interest). The parameter $\epsilon$ can be represented as:
$\epsilon = \phi(q)$,
where $\phi$ is a function that maps the context of the query $q$ to an appropriate numerical value.

Once the geometries $G_r$, $G_t$, functions $\mathcal{F}_s$, and parameters $\epsilon$ are delineated, the system constructs the precise spatial query $Q_s$. This query can be formally expressed by Equation \ref{eq:sql},
which ensures exact retrievals from the spatial database, maintaining both accuracy and relevance in the results. By leveraging these mathematical formulations, the system effectively translates spatial reasoning tasks into executable queries, facilitating robust spatial intelligence within the LLM framework.

\subsection{Hybrid spatial relevance scoring}\label{sec:ranking}
The spatial relevance score $f_s$ consists of two components: a score derived from 
sparse spatial retrieval from the spatial database
and a score from  dense spatial retrieval based on text similarity between the question and the spatial descriptions of candidate objects. 
Formally, we define:
\begin{equation}
f_s = \lambda_p f_s^\text{sparse} + \lambda_d f_s^\text{dense},
\end{equation}
where $\lambda_p$ and $\lambda_d$ are weighting coefficients controlling the contribution of each score.
\subsubsection{Sparse spatial relevance scoring}
Sparse spatial relevance is computed directly from the spatial database using explicit spatial relationships. The score is determined by the spatial query function $\mathcal{F}_s$, which computes the distance between reference and target objects. Formally, we define:
\begin{equation}
f_s^\text{sparse} = \begin{cases}
\displaystyle \frac{1}{1 + d(g_r, g_t)}, & \text{if } g_r \cap g_t = \emptyset \, \\
1, & \text{if } g_r \cap g_t \neq \emptyset \,
\end{cases}
\end{equation}
where $g_r$ and $g_t$ are reference and target spatial objects, respectively. $d(g_r, g_t)$ is a distance function measuring proximity in the spatial database. If $g_t$ overlaps with $g_r$, we assign a perfect relevance score of 1.
This ensures that objects within a region are maximally relevant, while those outside the region receive scores that decay with increasing distance.
\subsubsection{Dense spatial relevance scoring}
Unlike sparse scoring, dense spatial relevance is inferred from textual descriptions associated with spatial objects. We leverage an LLM to extract key spatial attributes from user queries and compare them with the descriptions of candidate objects.

\paragraph{Extracting spatial requirements and ranking via cosine similarity}
Given a user query $q$ and a set of text descriptions $d_t$ for spatial objects $G_t$, we extract the spatial requirements via an attention-based masking function:
\begin{equation}
v_{q,s} = \mathcal{E}(\mathcal{M}_s(q)), \quad v_{t,s} =\mathcal{E}(\mathcal{M}_s(d_t)),
\end{equation}
where $v_{q,s}$ and $v_{t,s}$ are dense vector representations of spatial features from query $q$ and text descriptions $d_t$. $\mathcal{M}_s$ is a function mapping input text to a spatial related text. $\mathcal{E}$ is the text encoder. The dense spatial relevance score $f_s^\text{dense}$ is computed via cosine similarity between two vectors.


\subsubsection{Hybrid spatial scoring as a generalized model}

We can demonstrate that hybrid spatial scoring generalizes both sparse and dense approaches:
\begin{itemize}[leftmargin=*]
\setlength\itemsep{-0.2em}
\item \textbf{Sparse-only case:} If $\lambda_d = 0$ , then
$f_s = \lambda_p f_s^\text{sparse}$,
reducing to a purely distance-based ranking.

\item \textbf{Dense-only case:} If $\lambda_p = 0$ , then
$f_s = \lambda_d f_s^\text{dense}$,
reducing to a purely semantic-based ranking.

\item \textbf{Hybrid case (General):} If both weights are nonzero, hybrid spatial scoring benefits from both explicit constraints and implicit relevance.
\end{itemize}

This formulation ensures that hybrid spatial scoring outperforms any single-scoring approach.







\subsection{Multi-objective generation}\label{sec:generation}
The semantic candidate set $C_k$ and the semantic relevance score $f_k$ are also calculated based on vector similarity, we put the details in Appendix \ref{app:dense_semantic}. After all the scores and candidate sets are acquired, the problem becomes a multi-objective optimization problem since each perspective (spatial and semantic) contributes from different aspects.

\subsubsection{Spatial-semantic Pareto frontier computation}

Given the spatial and semantic relevance scores, our goal is to identify the Pareto-optimal candidates that achieve the best trade-off between these objectives. A candidate $y$ is spatial-semantic Pareto-optimal if no other candidate dominates it in both spatial and semantic relevance.
Formally, the Pareto frontier $P(q)$ is defined as:
\begin{equation}
\begin{aligned}
P(q) = \{ &y \in C_s \cap C_k \ | \  \nexists \; y' \in C_s \cap C_k, \\
& f_s(q, y') \geq f_s(q, y) \ \text{and} \ f_k(q, y') \geq f_k(q, y),  \text{with at least one strict inequality} \}.
\end{aligned}
\end{equation}
This ensures that each candidate in $P(q)$ is non-dominated, meaning no other candidate is strictly better in both spatial and semantic relevance. 

\subsubsection{LLM-based trade-Off decision}

Once the Pareto frontier $P(q)$ is determined, we use an LLM to dynamically balance the trade-offs between spatial constraints and semantic preferences based on the context of the user query.
Specifically, the LLM receives the user query $q$, sparse spatial relevance scores $f_s^{\text{sparse}}$, and spatial object descriptions $d_y$ as input:
\begin{equation}
I = \left\{ q, \left( f_s^{\text{sparse}}(q, y) \ , d_y  \right) \ ,\forall y \in P(q)  \right\}.
\end{equation}
A dynamic weighting function $\lambda_s,\lambda_k=h(I)$ based on contextual information is extracted from the input, adjusting the importance of spatial vs. semantic relevance, where $h$ is a learned function capturing query-specific trade-offs.
The top-ranked candidate $y^*$ is selected by LLM: 
\begin{equation}
y^* = \arg\max_{y \in P(q)} \lambda_s^T f_s(q, y) + \lambda_k^T f_k(q, y),
\end{equation}
and the LLM generates a natural language response.
The system adapts to different query contexts instead of using a fixed weighting scheme. By structuring decision-making into discrete steps (candidate filtering $\rightarrow$ Pareto selection $\rightarrow$ trade-off balancing $\rightarrow$ response generation), the LLM avoids generating infeasible or illogical results.

\section{Experiment}
\subsection{Experiment setting}

\begin{wraptable}{r}{0.48\textwidth}
\vspace{-1em}
\caption{Overview of Datasets}
\label{tab:dataset_overview}
\centering
\resizebox{\linewidth}{!}{%
\begin{tabular}{lcc}
\hline
\textbf{Dataset} & \textbf{\#POIs} & \textbf{\#QA Pairs} \\
\hline
TourismQA-NYC & 9,470 & 17,448 \\
TourismQA-Miami & 2,640 & 133 \\
MapQA-ADJ & 92,415 & 50 \\
MapQA-AME & 92,415 & 231 \\
\hline
\end{tabular}
\vspace{-0.8cm}
}
\end{wraptable}

\paragraph{Datasets.}
We conduct our experiments on four datasets, including \textbf{TourismQA-NYC Dataset} and \textbf{TourismQA-Miami Dataset} datasets, which contains user questions about points of interest (POIs) in real-world cities. The questions are crawled from TripAdvisor posts, while the reviews of restaurants, attractions, and hotels are collected from travel forums and hotel booking websites.
For both datasets, we apply similar preprocessing steps, including removing POIs with missing review information and eliminating duplicate QA pairs.
\textbf{MapQA-Adjacent Dataset} (MapQA-ADJ) features questions such as “What [amenity] is adjacent to [location]?”, emphasizing topological relationships.
\textbf{MapQA-Amenities\_Around\_Specific Dataset} (MapQA-AME) focuses on proximity-based queries like “What are the [amenity] within 50m of [location]?”
The number of POIs and QA pairs for each dataset is summarized in Table~\ref{tab:dataset_overview}.

\vspace{-0.3cm}
\paragraph{Evaluation metrics.}
To assess the recommendation performance of different methods, we evaluate each model using five widely adopted metrics: Precision@k (The proportion of the top‑k returned items that are relevant to the query), Recall@k (The proportion of all relevant items for a query that appear within the top‑k results), F1@k (The harmonic mean of Precision@k and Recall@k), and NDCG@k (Normalized Discounted Cumulative Gain up to rank k), where $k\in\{1,3,5,10\}$. 


\begin{table*}[!ht]
\centering
\vspace{-0.3cm}
\caption{Performance comparison of models in TourismQA-NYC and TourismQA-Miami. }
\renewcommand{\arraystretch}{1.5}
\normalsize
\resizebox{\textwidth}{!}{%
  \begin{tabular}{@{}ll*{4}{*{4}{c}}@{}}
    \toprule
    \multirow{2}{*}{\textbf{Dataset}} &
    \multirow{2}{*}{\textbf{Method}} &
    \multicolumn{4}{c}{\textbf{Precision}} &
    \multicolumn{4}{c}{\textbf{Recall}} &
    \multicolumn{4}{c}{\textbf{F1}} &
    \multicolumn{4}{c}{\textbf{NDCG}} \\
    \cmidrule(lr){3-6} \cmidrule(lr){7-10}
    \cmidrule(lr){11-14} \cmidrule(lr){15-18} 
    & & @1 & @3 & @5 & @10
      & @1 & @3 & @5 & @10
      & @1 & @3 & @5 & @10
      & @1 & @3 & @5 & @10 \\
    \midrule
    \multirow{7}{*}{\textbf{TourismQA-Miami}}
   
      & \textbf{GeoLLM}              & 0.3158      & 0.2719      & 0.2368    & 0.2053      & 0.0365      & 0.0945     &0.1246     & 0.1824     & 0.0591      & 0.1094     & 0.1311       & 0.1623     
      & 0.3158     & 0.3200      & 0.3268     & 0.3600      \\
      & \textbf{Naive RAG}           & 0.2973      & 0.2252      & 0.2054      & 0.1973      &0.0814    & 0.1124      & 0.1491      & 0.2301     & 0.1005       &  0.1099      & 0.1310     & 0.1642    
      & 0.2973     & 0.2753      & 0.2849      & 0.3403     \\
      & \textbf{TE}        &0.3421      & 0.3070    & 0.2316     &0.1842        & 0.0567      &0.1544    & 0.1821     &0.2231       & 0.0781    &  0.1558   & 0.1515    & 0.1543     
      &0.3421    &0.3766     &0.3890    & 0.4748      \\
    
      & \textbf{SD}                  &0.1053     & 0.0351      &0.0211    &0.0105     & 0.0075      &  0.0075     &  0.0075     &  0.0075    &0.0138      & 0.0119      & 0.0105      & 0.0082     
      &  0.1053     & 0.1053     &  0.1053     &  0.1053      \\
      & \textbf{ST}                  & 0.3947      & 0.1316      &  0.0789      & 0.0395      & 0.0873     & 0.0873     & 0.0873     & 0.0873     &  0.1115      &0.0730      &0.0569      &0.0386      
      & 0.3947      & 0.3947   & 0.3947    & 0.3947     \\
         & \textbf{Spatial-RAG(GPT3.5-Turbo)}         & \underline{\textbf{0.5455}}      & \textbf{0.4141}       & \underline{\textbf{0.4000}}    & \underline{\textbf{0.3152}}     &\underline{\textbf{ 0.1081} }     & \underline{\textbf{0.2072}}      & \underline{\textbf{0.2972}}     & \underline{\textbf{0.3732}}     & \underline{\textbf{0.1540}}   &\underline{\textbf{0.2262 }}     &\underline{\textbf{0.2765}}        &\underline{\textbf{0.2817}}       
         & \underline{\textbf{0.5455}}      & \underline{\textbf{0.492}}      & \underline{\textbf{0.5276}}      & \underline{\textbf{0.5440}}     \\
      & \textbf{Spatial-RAG (GPT4-Turbo)} & \textbf{0.5152}      & \underline{\textbf{0.4242}}     & \textbf{0.3758}     & \textbf{0.2939}     & \textbf{0.1026}    &  \textbf{0.2037}       & \textbf{0.2841}      & \textbf{0.3565}      & \textbf{0.1454}    &\textbf{0.2246}       &\textbf{0.2619}      &\textbf{0.2656 }     
      &\textbf{0.5152}     & \textbf{0.4852}    & \textbf{0.4983}      & \textbf{0.5079}      \\
\midrule
\multirow{7}{*}{\textbf{TourismQA-NYC}}
  
      & \textbf{GeoLLM}              & 0.3650 &  0.3292 & 0.3020 & 0.2725 & 0.0168  & 0.0433  & 0.0614 & 0.1033    & 0.0311 & 0.0688 & 0.0907&0.1328 
      & 0.3650 & 0.3626 & 0.3708 & 0.4248 \\
      & \textbf{Naive RAG}           & \textbf{0.4923} & 0.4447 & 0.4245 & 0.4181 & 0.0214 &  0.0556 & 0.0889 & 0.1678 &  0.0399 & 0.0932&  0.1345 &\textbf{0.2167} 
      & \textbf{0.4923} &0.4551 & 0.4528 & 0.5058 \\
      & \textbf{TE}                  &  0.2250   & 0.2433     & 0.2395    & 0.2325     & 0.0101       &  0.0330    & 0.0506    &0.0934     &0.0189    &  0.0541      &    0.0767   &  0.1199     
      & 0.2250   & 0.2873      & 0.3533   & 0.5142   \\
      
      & \textbf{SD}                  &   0.2425   & 0.2492     &0.2455     & 0.2402     & 0.0103      & 0.0314     & 0.0520      &0.0945     & 0.0193      & 0.0520      & 0.0758      & 0.1194      
      & 0.2425      & 0.2694      & 0.2926      & 0.3624      \\
          & \textbf{ST}                  & 0.4725     &  \textbf{0.4608}     &  \textbf{0.4460}     & \underline{\textbf{0.4323}}      &\textbf{0.0238}     & \underline{\textbf{ 0.0644}}     & \underline{\textbf{0.1026}}     & \underline{\textbf{0.1846}}       & \textbf{0.0433}      & \underline{\textbf{0.1026}}     &\underline{\textbf{0.1475}}       &\underline{\textbf{0.2281}}     
          & 0.4725     & \textbf{0.4686}      & \textbf{0.4745}      & \textbf{0.5296}      \\
      
      & \textbf{Spatial-RAG (GPT3.5-Turbo)}          & 0.4611 &0.4352 & 0.4144 &0.4098 & 0.0188 &  0.0485 & 0.0751 & 0.1493 & 0.0350 &0.0828 & 0.1188 &0.2000 
      & 0.4611 & 0.4545 &0.4534 &0.5097 \\
      & \textbf{Spatial-RAG (GPT4-Turbo)} & \underline{\textbf{0.5665}} & \underline{\textbf{0.4875 }} & \underline{\textbf{0.4555}} & \textbf{0.4251} & \underline{\textbf{0.0274}} & \textbf{0.0611} & \textbf{0.0908} & 
      \textbf{0.1691} & \underline{\textbf{ 0.0483}} & \textbf{0.0996} &\textbf{0.1384}  & 0.2153 
      & \underline{\textbf{0.5665}} & \underline{\textbf{0.5174}} & \underline{\textbf{0.5065}} & \underline{\textbf{ 0.5574}} \\
    \bottomrule
  \end{tabular}%
}
\label{tab:main1}
\vspace{-0.1cm}
\end{table*}

\begin{table*}[!ht]
\centering
\caption{Performance comparison of models on MapQA-ADJ and MapQA-AME.}
\renewcommand{\arraystretch}{1.2}
\resizebox{\textwidth}{!}{%
\begin{tabular}{llccccccc}
\toprule
\textbf{Dataset} & \textbf{Metric} & \multicolumn{7}{c}{\textbf{Methods}} \\
\cmidrule(lr){3-9}
 & & SD & TE & ST & Naive RAG & GeoLLM & Spatial-RAG (GPT-3.5-Turbo) & Spatial-RAG (GPT-4.0-Turbo) \\
\midrule
\multirow{4}{*}{\textbf{MapQA-ADJ}}
 & Precision & 0.4600 & 0.4200 & \underline{\textbf{0.5600}} & 0.3455 & 0.4558 & \textbf{0.5467} & 0.5057 \\
 & Recall    & 0.4200 & 0.3750 & 0.5150         & 0.6170 & \textbf{0.6685} & 0.5350                   & \underline{\textbf{0.7819}} \\
 & F1        & 0.4333 & 0.3880 & \textbf{0.5280} & 0.4067 & 0.5073 & 0.5247                   & \underline{\textbf{0.5625}} \\
 & NDCG      & 0.4600 & 0.4200 & 0.5600 &  0.5866 & \textbf{0.6132} & 0.5800 &  \underline{\textbf{0.7391}} \\
\midrule
\multirow{4}{*}{\textbf{MapQA-AME}}
 & Precision & 0.4323 & 0.2751 & \textbf{0.5714} & 0.2452 & 0.3729 & 0.5566                   & \underline{\textbf{0.6645}} \\
 & Recall    & 0.3923 & 0.2351 & 0.5173         & 0.6075 & \textbf{0.8333} & 0.7446                   & \underline{\textbf{0.9072}} \\
 & F1        & 0.4054 & 0.2482 & 0.5346         & 0.3263 & 0.4744 & \textbf{0.6100}           & \underline{\textbf{0.7298}} \\
 & NDCG      &  0.4323 & 0.2751 & 0.5714         & 0.6018 & \textbf{0.8199} & 0.7264                  & \underline{\textbf{0.8719}} \\
\bottomrule
\end{tabular}%
}
\label{tab:main2}
\end{table*}

\begin{table*}[!ht]
\centering
\vspace{-0.3cm}
\caption{Ablation results.}
\renewcommand{\arraystretch}{1.3}
\resizebox{\textwidth}{!}{%
\begin{tabular}{lcccccccccccccccc}
\toprule
\textbf{Method} &
\multicolumn{4}{c}{\textbf{Precision}} &
\multicolumn{4}{c}{\textbf{Recall}} &
\multicolumn{4}{c}{\textbf{F1}} &
\multicolumn{4}{c}{\textbf{NDCG}} \\
\cmidrule(lr){2-5} \cmidrule(lr){6-9} \cmidrule(lr){10-13} \cmidrule(lr){14-17} 
& @1 & @3 & @5 & @10 &
  @1 & @3 & @5 & @10 &
  @1 & @3 & @5 & @10 &
  @1 & @3 & @5 & @10 \\
\midrule
Spatial-RAG(GPT4-Turbo)           & 0.5665 & 0.4875 & 0.4555 & 0.4251 & 0.0274 & 0.0611 & 0.0908 & 0.1691 & 0.0483 & 0.0996 & 0.1384 & 0.2153 
& 0.5665 & 0.5174 & 0.5065 & 0.5574\\
\midrule
w/o sparse spatial     & 0.3807 & 0.3545& 0.3503 & 0.3307 & 0.0147 & 0.0431 & 0.0701 & 0.1305 & 0.0276 & 0.0701 &0.1058 & 0.1667 
& 0.3807 & 0.3608 &0.3570 & 0.3455 \\
w/o dense spatial      & 0.5569 & 0.4954 & 0.4628 & 0.4295 & 0.0263 & 0.0606 & 0.0919 & 0.1679 & 0.0466 & 0.0986 & 0.1392 & 0.2132 
& 0.5569 & 0.5240 &0.5132 & 0.5587 \\
w/o dense semantic     & 0.4986 & 0.4311 & 0.4165 & 0.3806 & 0.0240 & 0.0553 & 0.0844 & 0.1499 & 0.0421 & 0.0883 & 0.1262 & 0.1896 
& 0.4986 & 0.4679 & 0.4739 & 0.5219 \\
Scratch                & 0.5392 & 0.4935 & 0.4706 & 0.4343 & 0.0236 & 0.0535 & 0.0829 & 0.1574 & 0.0435 & 0.0916 & 0.1323 & 0.2105 
& 0.5392 & 0.5105 & 0.5042 & 0.5374 \\
w/o RAG                & 0.3584 & 0.3333 & 0.3028 & 0.2644 & 0.0121 & 0.0323 & 0.0484 & 0.0846 & 0.0231 & 0.0568 & 0.0796 & 0.1191 
& 0.3584 & 0.3570 & 0.3653 & 0.4186 \\
\bottomrule
\end{tabular}%
}
\vspace{-0.2cm}
\label{tab:ablation}
\end{table*}

\paragraph{Models for comparison.}
We used GPT-3.5-Turbo and GPT-4-Turbo as the LLM in our framework and compared them against the following methods:
\textbf{GeoLLM} \cite{manvigeollm}encodes the spatial objects to address and enrich their context by adding spatial information of nearby spatial objects. 
\textbf{Naive RAG} \cite{lewis2020retrieval} saves all spatial objects' descriptions in a vector database and retrieves the most relevant objects based on vector similarity.
\textbf{Text embedding} (TE) \cite{cakaloglu2020text} is a greedy method minimizing the distance between the vector embeddings of the text description of the reference object and the target object.
\textbf{Sort-by-distance} (SD) \cite{contractor2021joint} ranks the candidate spatial objects based on their distance to the reference objects in the spatial question.
\textbf{Spatial-text} (ST) computes the embeddings of the user’s question and compares the similarity between the question embedding and the text description embedding of the target object. Additionally, the object’s location is encoded as a distance score. The answer is determined based on the average of these scores.

\subsection{Spatial-RAG vs. Baselines}
The comparative performance results across methods and datasets are summarized in Tables~\ref{tab:main1} and~\ref{tab:main2}. Our Spatial-RAG (GPT4-Turbo) consistently surpasses baselines, including Naive RAG, GeoLLM, and template-based methods (ST, TE, SD). Specifically, a 19.9\% improvement in Precision@1 was achieved by our Spatial-RAG (GPT-4-Turbo) compared to the best baseline model (ST) on TourismQA-NYC. It improved NDCG by 32.0\% on MapQA-ADJ and by 52.6\% on MapQA-AME compared to ST. On the MapQA-AME dataset, it notably increased Recall by 75.4\% over ST. Compared to GeoLLM, Spatial-RAG (GPT-4-Turbo) achieved 95.4\% higher Recall@10 on TourismQA-Miami. 
This advantage arises primarily from Spatial-RAG’s structured spatial retrieval pipeline, which leverages a spatial database to accurately retrieve relevant candidates that satisfy the spatial constraints, a capability lacking in embedding-based methods like Naive RAG and purely textual or parametric models (GeoLLM, SD, TE). Meanwhile, although ST occasionally leads in single-relationship precision (e.g., Precision@1 on MapQA-ADJ), Spatial-RAG’s GPT4-powered re-ranking provides superior overall performance through nuanced multi-constraint spatial reasoning. Among evaluation metrics, Spatial-RAG sees greater gains in Recall and NDCG than in Precision, highlighting that structured spatial queries significantly enhance completeness of retrieved results and the accuracy of ranking order. Finally, Spatial-RAG (GPT4-Turbo) consistently outperforms its GPT3.5 counterpart, underscoring GPT4’s advanced reasoning capabilities and improved candidate-ranking accuracy.

Replacing GPT3.5-Turbo with GPT4-Turbo within the identical retrieval pipeline contributes a further +2 pp on average, indicating that model capacity and spatially constrained retrieval are additive rather than redundant. This improvement is especially evident on structured spatial QA tasks such as those in MapQA, which feature well-defined relations like “adjacent” and “within 50m.” In contrast, the performance gap is narrower on noisier, open-domain datasets like TourismQA, suggesting that GPT4’s capacity brings greater benefit when precise spatial reasoning is required.
The above results establish Spatial-RAG as the new state of the art, delivering statistically robust, across-the-board improvements relative to all prior baselines.

\subsection{Ablation study}
To validate our technical contributions, we conducted five ablation studies. First, we examined the impact of removing individual modules: the sparse spatial module (w/o sparse spatial), dense spatial module (w/o dense spatial), and dense semantic module (w/o dense semantic). Next, we evaluated two system-level variations: generating SQL from scratch (Scratch) and removing the RAG component (w/o RAG).

Table \ref{tab:ablation} summarizes the ablation results. Removing any component leads to a performance decline, highlighting the importance of each module. Notably, excluding the sparse spatial component results in the most significant drop, underscoring the value of integrating the spatial database with the LLM. The Scratch setting shows only a slight decrease, suggesting that GPT4-Turbo can internally formulate SQL queries when given essential inputs, even without templates. The w/o RAG variant performs worst across all metrics, indicating that retrieval-augmented generation is fundamental to the system's overall effectiveness.


\subsection{Case studies}
\begin{figure*}[t!]
  \centering
  \includegraphics[width=\textwidth]{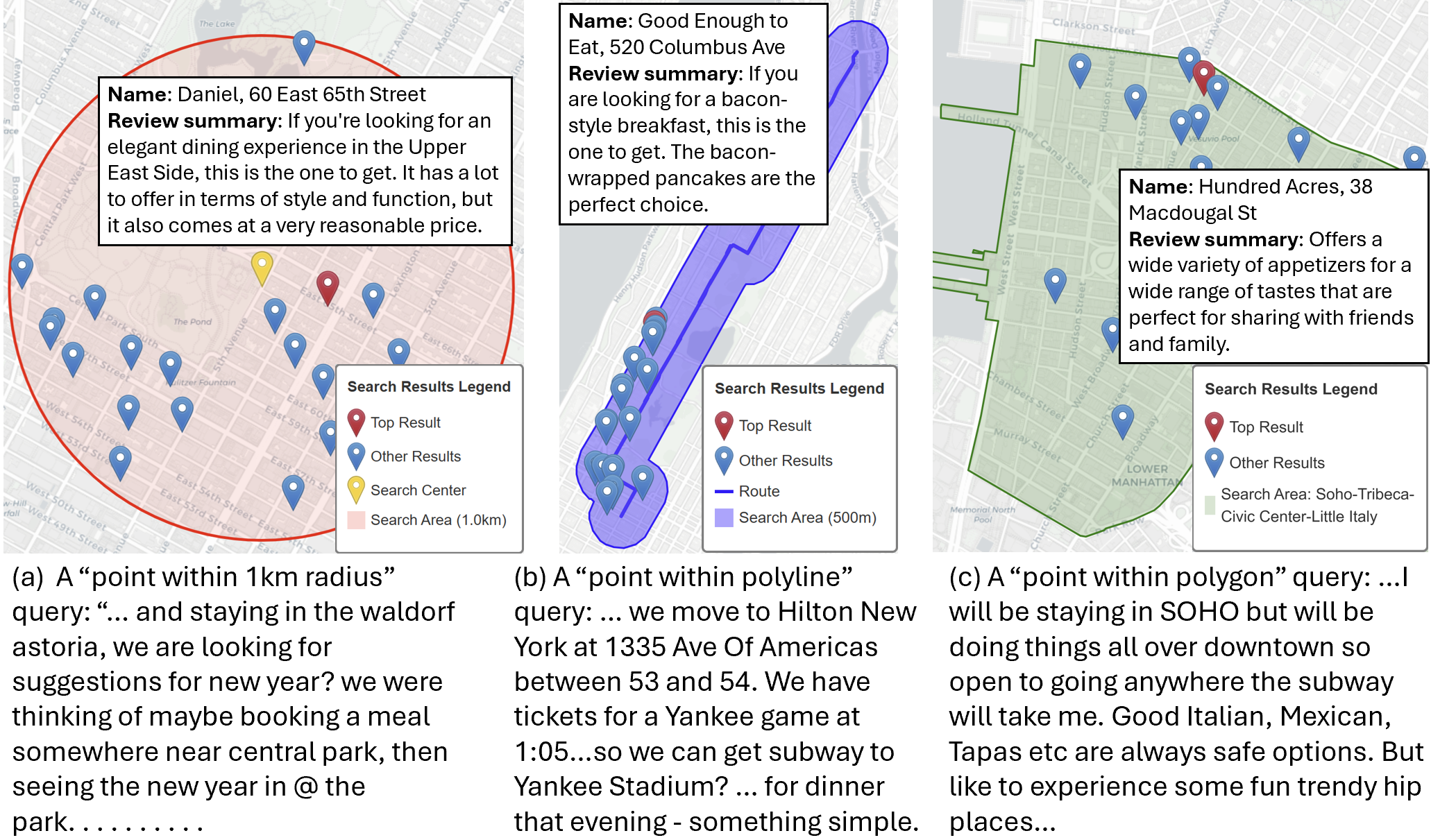}
  \vspace{-0.5cm}
  \caption{Three POI queries with different reference objects.}
  \vspace{-0.5cm}
  \label{fig:3example}
\end{figure*}
The visualizations of three POI queries with different reference objects are given in Figure \ref{fig:3example}. Spatial-RAG effectively identifies the user’s intent within noisy queries, detecting that the user needs to travel near a location, along a route, or within a region. Based on this, Spatial-RAG applies proper spatial constraints to filter spatial objects. In contrast, traditional methods, such as distance-based approaches, always generate a buffer zone around a single location, which may not necessarily include POIs along the user’s travel route. Our framework provides a more context-aware understanding of the user’s spatial intent, enabling more precise and relevant recommendations based on both location and user preferences.



\section{Conclusion}
Spatial-RAG enhances LLMs' spatial reasoning by integrating structured spatial retrieval with natural language understanding, bridging the gap between spatial databases and LLM-driven question answering.  Extensive evaluations show that Spatial-RAG outperforms existing methods, highlighting its potential to advance spatial analysis, tourism recommendation, and geographic QA. 

\begin{ack}
Use unnumbered first level headings for the acknowledgments. All acknowledgments
go at the end of the paper before the list of references. Moreover, you are required to declare
funding (financial activities supporting the submitted work) and competing interests (related financial activities outside the submitted work).
More information about this disclosure can be found at: \url{https://neurips.cc/Conferences/2025/PaperInformation/FundingDisclosure}.

Do {\bf not} include this section in the anonymized submission, only in the final paper. You can use the \texttt{ack} environment provided in the style file to automatically hide this section in the anonymized submission.
\end{ack}

\bibliographystyle{plain}
\bibliography{example_paper}

\begin{thebibliography}{10}

\bibitem{cakaloglu2020text}
Tolgahan Cakaloglu, Christian Szegedy, and Xiaowei Xu.
\newblock Text embeddings for retrieval from a large knowledge base.
\newblock In {\em International Conference on Research Challenges in Information Science}, pages 338--351. Springer, 2020.

\bibitem{chen2014parameterized}
Wei Chen.
\newblock Parameterized spatial sql translation for geographic question answering.
\newblock In {\em 2014 IEEE international conference on semantic computing}, pages 23--27. IEEE, 2014.

\bibitem{christmann2024rag}
Philipp Christmann and Gerhard Weikum.
\newblock Rag-based question answering over heterogeneous data and text.
\newblock {\em arXiv preprint arXiv:2412.07420}, 2024.

\bibitem{contractor2021joint}
Danish Contractor, Shashank Goel, Mausam, and Parag Singla.
\newblock Joint spatio-textual reasoning for answering tourism questions.
\newblock In {\em Proceedings of the Web Conference 2021}, pages 1978--1989, 2021.

\bibitem{faisal2023geographic}
Fahim Faisal and Antonios Anastasopoulos.
\newblock Geographic and geopolitical biases of language models.
\newblock In {\em Proceedings of the 3rd Workshop on Multi-lingual Representation Learning (MRL)}, pages 139--163, 2023.

\bibitem{fan2024survey}
Wenqi Fan, Yujuan Ding, Liangbo Ning, Shijie Wang, Hengyun Li, Dawei Yin, Tat-Seng Chua, and Qing Li.
\newblock A survey on rag meeting llms: Towards retrieval-augmented large language models.
\newblock In {\em Proceedings of the 30th ACM SIGKDD Conference on Knowledge Discovery and Data Mining}, pages 6491--6501, 2024.

\bibitem{gurnee2024language}
Wes Gurnee and Max Tegmark.
\newblock Language models represent space and time.
\newblock In {\em The Twelfth International Conference on Learning Representations}, 2024.

\bibitem{he2024g}
Xiaoxin He, Yijun Tian, Yifei Sun, Nitesh~V Chawla, Thomas Laurent, Yann LeCun, Xavier Bresson, and Bryan Hooi.
\newblock G-retriever: Retrieval-augmented generation for textual graph understanding and question answering.
\newblock {\em arXiv preprint arXiv:2402.07630}, 2024.

\bibitem{ji2023evaluating}
Yuhan Ji and Song Gao.
\newblock Evaluating the effectiveness of large language models in representing textual descriptions of geometry and spatial relations (short paper).
\newblock In {\em 12th International Conference on Geographic Information Science (GIScience 2023)}, pages 43--1. Schloss Dagstuhl--Leibniz-Zentrum f{\"u}r Informatik, 2023.

\bibitem{kefalidis2024question}
Sergios-Anestis Kefalidis, Dharmen Punjani, Eleni Tsalapati, Konstantinos Plas, Maria-Aggeliki Pollali, Pierre Maret, and Manolis Koubarakis.
\newblock The question answering system geoqa2 and a new benchmark for its evaluation.
\newblock {\em International Journal of Applied Earth Observation and Geoinformation}, 134:104203, 2024.

\bibitem{lewis2020retrieval}
Patrick Lewis, Ethan Perez, Aleksandra Piktus, Fabio Petroni, Vladimir Karpukhin, Naman Goyal, Heinrich K{\"u}ttler, Mike Lewis, Wen-tau Yih, Tim Rockt{\"a}schel, et~al.
\newblock Retrieval-augmented generation for knowledge-intensive nlp tasks.
\newblock {\em Advances in neural information processing systems}, 33:9459--9474, 2020.

\bibitem{li2024advancing}
Fangjun Li, David~C Hogg, and Anthony~G Cohn.
\newblock Advancing spatial reasoning in large language models: An in-depth evaluation and enhancement using the stepgame benchmark.
\newblock In {\em Proceedings of the AAAI Conference on Artificial Intelligence}, volume~38, pages 18500--18507, 2024.

\bibitem{li2023geolm}
Zekun Li, Wenxuan Zhou, Yao-Yi Chiang, and Muhao Chen.
\newblock Geolm: Empowering language models for geospatially grounded language understanding.
\newblock {\em arXiv preprint arXiv:2310.14478}, 2023.

\bibitem{lietard2021language}
Bastien Li{\'e}tard, Mostafa Abdou, and Anders S{\o}gaard.
\newblock Do language models know the way to rome?
\newblock {\em arXiv preprint arXiv:2109.07971}, 2021.

\bibitem{mai2024opportunities}
Gengchen Mai, Weiming Huang, Jin Sun, Suhang Song, Deepak Mishra, Ninghao Liu, Song Gao, Tianming Liu, Gao Cong, Yingjie Hu, et~al.
\newblock On the opportunities and challenges of foundation models for geoai (vision paper).
\newblock {\em ACM Transactions on Spatial Algorithms and Systems}, 2024.

\bibitem{mai2021geographic}
Gengchen Mai, Krzysztof Janowicz, Rui Zhu, Ling Cai, and Ni~Lao.
\newblock Geographic question answering: challenges, uniqueness, classification, and future directions.
\newblock {\em AGILE: GIScience series}, 2:8, 2021.

\bibitem{manvi2024large}
Rohin Manvi, Samar Khanna, Marshall Burke, David~B Lobell, and Stefano Ermon.
\newblock Large language models are geographically biased.
\newblock In {\em International Conference on Machine Learning}, pages 34654--34669. PMLR, 2024.

\bibitem{manvigeollm}
Rohin Manvi, Samar Khanna, Gengchen Mai, Marshall Burke, David~B Lobell, and Stefano Ermon.
\newblock Geollm: Extracting geospatial knowledge from large language models.
\newblock In {\em The Twelfth International Conference on Learning Representations}, 2024.

\bibitem{roberts2023gpt4geo}
Jonathan Roberts, Timo L{\"u}ddecke, Sowmen Das, Kai Han, and Samuel Albanie.
\newblock Gpt4geo: How a language model sees the world's geography.
\newblock {\em arXiv preprint arXiv:2306.00020}, 2023.

\bibitem{siriwardhana2023improving}
Shamane Siriwardhana, Rivindu Weerasekera, Elliott Wen, Tharindu Kaluarachchi, Rajib Rana, and Suranga Nanayakkara.
\newblock Improving the domain adaptation of retrieval augmented generation (rag) models for open domain question answering.
\newblock {\em Transactions of the Association for Computational Linguistics}, 11:1--17, 2023.

\bibitem{wu2024torchspatial}
Nemin Wu, Qian Cao, Zhangyu Wang, Zeping Liu, Yanlin Qi, Jielu Zhang, Joshua Ni, Xiaobai Yao, Hongxu Ma, Lan Mu, et~al.
\newblock Torchspatial: A location encoding framework and benchmark for spatial representation learning.
\newblock {\em arXiv preprint arXiv:2406.15658}, 2024.

\bibitem{xietravelplanner}
Jian Xie, Kai Zhang, Jiangjie Chen, Tinghui Zhu, Renze Lou, Yuandong Tian, Yanghua Xiao, and Yu~Su.
\newblock Travelplanner: A benchmark for real-world planning with language agents.
\newblock In {\em Forty-first International Conference on Machine Learning}, 2024.

\bibitem{yu2024llms}
Jiong Yu, Sixing Wu, Jiahao Chen, and Wei Zhou.
\newblock Llms as collaborator: Demands-guided collaborative retrieval-augmented generation for commonsense knowledge-grounded open-domain dialogue systems.
\newblock In {\em Findings of the Association for Computational Linguistics: EMNLP 2024}, pages 13586--13612, 2024.

\bibitem{zhang2024bb}
Yifan Zhang, Zhiyun Wang, Zhengting He, Jingxuan Li, Gengchen Mai, Jianfeng Lin, Cheng Wei, and Wenhao Yu.
\newblock Bb-geogpt: A framework for learning a large language model for geographic information science.
\newblock {\em Information Processing \& Management}, 61(5):103808, 2024.

\end{thebibliography}


\newpage
\section*{NeurIPS Paper Checklist}

\begin{enumerate}

\item {\bf Claims}
    \item[] Question: Do the main claims made in the abstract and introduction accurately reflect the paper's contributions and scope?
    \item[] Answer: \answerYes{} 
    \item[] Justification: {the main claims made in the abstract and introduction accurately reflect the paper's contributions and scope}
    \item[] Guidelines:
    \begin{itemize}
        \item The answer NA means that the abstract and introduction do not include the claims made in the paper.
        \item The abstract and/or introduction should clearly state the claims made, including the contributions made in the paper and important assumptions and limitations. A No or NA answer to this question will not be perceived well by the reviewers. 
        \item The claims made should match theoretical and experimental results, and reflect how much the results can be expected to generalize to other settings. 
        \item It is fine to include aspirational goals as motivation as long as it is clear that these goals are not attained by the paper. 
    \end{itemize}

\item {\bf Limitations}
    \item[] Question: Does the paper discuss the limitations of the work performed by the authors?
    \item[] Answer: \answerYes{} 
    \item[] Justification: We discuss the limitations of the work in appendix E
    \item[] Guidelines:
    \begin{itemize}
        \item The answer NA means that the paper has no limitation while the answer No means that the paper has limitations, but those are not discussed in the paper. 
        \item The authors are encouraged to create a separate "Limitations" section in their paper.
        \item The paper should point out any strong assumptions and how robust the results are to violations of these assumptions (e.g., independence assumptions, noiseless settings, model well-specification, asymptotic approximations only holding locally). The authors should reflect on how these assumptions might be violated in practice and what the implications would be.
        \item The authors should reflect on the scope of the claims made, e.g., if the approach was only tested on a few datasets or with a few runs. In general, empirical results often depend on implicit assumptions, which should be articulated.
        \item The authors should reflect on the factors that influence the performance of the approach. For example, a facial recognition algorithm may perform poorly when image resolution is low or images are taken in low lighting. Or a speech-to-text system might not be used reliably to provide closed captions for online lectures because it fails to handle technical jargon.
        \item The authors should discuss the computational efficiency of the proposed algorithms and how they scale with dataset size.
        \item If applicable, the authors should discuss possible limitations of their approach to address problems of privacy and fairness.
        \item While the authors might fear that complete honesty about limitations might be used by reviewers as grounds for rejection, a worse outcome might be that reviewers discover limitations that aren't acknowledged in the paper. The authors should use their best judgment and recognize that individual actions in favor of transparency play an important role in developing norms that preserve the integrity of the community. Reviewers will be specifically instructed to not penalize honesty concerning limitations.
    \end{itemize}

\item {\bf Theory assumptions and proofs}
    \item[] Question: For each theoretical result, does the paper provide the full set of assumptions and a complete (and correct) proof?
    \item[] Answer: \answerNA{} 
    \item[] Justification: This paper does not involves theoretical result.
    \item[] Guidelines:
    \begin{itemize}
        \item The answer NA means that the paper does not include theoretical results. 
        \item All the theorems, formulas, and proofs in the paper should be numbered and cross-referenced.
        \item All assumptions should be clearly stated or referenced in the statement of any theorems.
        \item The proofs can either appear in the main paper or the supplemental material, but if they appear in the supplemental material, the authors are encouraged to provide a short proof sketch to provide intuition. 
        \item Inversely, any informal proof provided in the core of the paper should be complemented by formal proofs provided in appendix or supplemental material.
        \item Theorems and Lemmas that the proof relies upon should be properly referenced. 
    \end{itemize}

    \item {\bf Experimental result reproducibility}
    \item[] Question: Does the paper fully disclose all the information needed to reproduce the main experimental results of the paper to the extent that it affects the main claims and/or conclusions of the paper (regardless of whether the code and data are provided or not)?
    \item[] Answer: \answerYes{} 
    \item[] Justification: We propose a framework and describe the architecture clearly and fully. 
    \item[] Guidelines:
    \begin{itemize}
        \item The answer NA means that the paper does not include experiments.
        \item If the paper includes experiments, a No answer to this question will not be perceived well by the reviewers: Making the paper reproducible is important, regardless of whether the code and data are provided or not.
        \item If the contribution is a dataset and/or model, the authors should describe the steps taken to make their results reproducible or verifiable. 
        \item Depending on the contribution, reproducibility can be accomplished in various ways. For example, if the contribution is a novel architecture, describing the architecture fully might suffice, or if the contribution is a specific model and empirical evaluation, it may be necessary to either make it possible for others to replicate the model with the same dataset, or provide access to the model. In general. releasing code and data is often one good way to accomplish this, but reproducibility can also be provided via detailed instructions for how to replicate the results, access to a hosted model (e.g., in the case of a large language model), releasing of a model checkpoint, or other means that are appropriate to the research performed.
        \item While NeurIPS does not require releasing code, the conference does require all submissions to provide some reasonable avenue for reproducibility, which may depend on the nature of the contribution. For example
        \begin{enumerate}
            \item If the contribution is primarily a new algorithm, the paper should make it clear how to reproduce that algorithm.
            \item If the contribution is primarily a new model architecture, the paper should describe the architecture clearly and fully.
            \item If the contribution is a new model (e.g., a large language model), then there should either be a way to access this model for reproducing the results or a way to reproduce the model (e.g., with an open-source dataset or instructions for how to construct the dataset).
            \item We recognize that reproducibility may be tricky in some cases, in which case authors are welcome to describe the particular way they provide for reproducibility. In the case of closed-source models, it may be that access to the model is limited in some way (e.g., to registered users), but it should be possible for other researchers to have some path to reproducing or verifying the results.
        \end{enumerate}
    \end{itemize}

\item {\bf Open access to data and code}
    \item[] Question: Does the paper provide open access to the data and code, with sufficient instructions to faithfully reproduce the main experimental results, as described in supplemental material?
    \item[] Answer: \answerYes{} 
    \item[] Justification: We will release of code and data once get accepted.
    \item[] Guidelines:
    \begin{itemize}
        \item The answer NA means that paper does not include experiments requiring code.
        \item Please see the NeurIPS code and data submission guidelines (\url{https://nips.cc/public/guides/CodeSubmissionPolicy}) for more details.
        \item While we encourage the release of code and data, we understand that this might not be possible, so “No” is an acceptable answer. Papers cannot be rejected simply for not including code, unless this is central to the contribution (e.g., for a new open-source benchmark).
        \item The instructions should contain the exact command and environment needed to run to reproduce the results. See the NeurIPS code and data submission guidelines (\url{https://nips.cc/public/guides/CodeSubmissionPolicy}) for more details.
        \item The authors should provide instructions on data access and preparation, including how to access the raw data, preprocessed data, intermediate data, and generated data, etc.
        \item The authors should provide scripts to reproduce all experimental results for the new proposed method and baselines. If only a subset of experiments are reproducible, they should state which ones are omitted from the script and why.
        \item At submission time, to preserve anonymity, the authors should release anonymized versions (if applicable).
        \item Providing as much information as possible in supplemental material (appended to the paper) is recommended, but including URLs to data and code is permitted.
    \end{itemize}

\item {\bf Experimental setting/details}
    \item[] Question: Does the paper specify all the training and test details (e.g., data splits, hyperparameters, how they were chosen, type of optimizer, etc.) necessary to understand the results?
    \item[] Answer: \answerYes{} 
    \item[] Justification: We specify all the training and test details in the main text and appendix.
    \item[] Guidelines:
    \begin{itemize}
        \item The answer NA means that the paper does not include experiments.
        \item The experimental setting should be presented in the core of the paper to a level of detail that is necessary to appreciate the results and make sense of them.
        \item The full details can be provided either with the code, in appendix, or as supplemental material.
    \end{itemize}

\item {\bf Experiment statistical significance}
    \item[] Question: Does the paper report error bars suitably and correctly defined or other appropriate information about the statistical significance of the experiments?
    \item[] Answer: \answerNo{} 
    \item[] Justification:  We follow the convention in prior works.
    \item[] Guidelines:
    \begin{itemize}
        \item The answer NA means that the paper does not include experiments.
        \item The authors should answer "Yes" if the results are accompanied by error bars, confidence intervals, or statistical significance tests, at least for the experiments that support the main claims of the paper.
        \item The factors of variability that the error bars are capturing should be clearly stated (for example, train/test split, initialization, random drawing of some parameter, or overall run with given experimental conditions).
        \item The method for calculating the error bars should be explained (closed form formula, call to a library function, bootstrap, etc.)
        \item The assumptions made should be given (e.g., Normally distributed errors).
        \item It should be clear whether the error bar is the standard deviation or the standard error of the mean.
        \item It is OK to report 1-sigma error bars, but one should state it. The authors should preferably report a 2-sigma error bar than state that they have a 96\% CI, if the hypothesis of Normality of errors is not verified.
        \item For asymmetric distributions, the authors should be careful not to show in tables or figures symmetric error bars that would yield results that are out of range (e.g. negative error rates).
        \item If error bars are reported in tables or plots, The authors should explain in the text how they were calculated and reference the corresponding figures or tables in the text.
    \end{itemize}

\item {\bf Experiments compute resources}
    \item[] Question: For each experiment, does the paper provide sufficient information on the computer resources (type of compute workers, memory, time of execution) needed to reproduce the experiments?
    \item[] Answer: \answerYes{} 
    \item[] Justification: We provided sufficient information on the computer resources in the main text and appendix.
    \begin{itemize}
        \item The answer NA means that the paper does not include experiments.
        \item The paper should indicate the type of compute workers CPU or GPU, internal cluster, or cloud provider, including relevant memory and storage.
        \item The paper should provide the amount of compute required for each of the individual experimental runs as well as estimate the total compute. 
        \item The paper should disclose whether the full research project required more compute than the experiments reported in the paper (e.g., preliminary or failed experiments that didn't make it into the paper). 
    \end{itemize}
    
\item {\bf Code of ethics}
    \item[] Question: Does the research conducted in the paper conform, in every respect, with the NeurIPS Code of Ethics \url{https://neurips.cc/public/EthicsGuidelines}?
    \item[] Answer: \answerYes{} 
    \item[] Justification:  the research conducted in the paper conformed, in every respect, with the NeurIPS Code of Ethics.
    \item[] Guidelines:
    \begin{itemize}
        \item The answer NA means that the authors have not reviewed the NeurIPS Code of Ethics.
        \item If the authors answer No, they should explain the special circumstances that require a deviation from the Code of Ethics.
        \item The authors should make sure to preserve anonymity (e.g., if there is a special consideration due to laws or regulations in their jurisdiction).
    \end{itemize}

\item {\bf Broader impacts}
    \item[] Question: Does the paper discuss both potential positive societal impacts and negative societal impacts of the work performed?
    \item[] Answer: \answerNA{} 
    \item[] Justification: there is no societal impact of the work performed
    \item[] Guidelines:
    \begin{itemize}
        \item The answer NA means that there is no societal impact of the work performed.
        \item If the authors answer NA or No, they should explain why their work has no societal impact or why the paper does not address societal impact.
        \item Examples of negative societal impacts include potential malicious or unintended uses (e.g., disinformation, generating fake profiles, surveillance), fairness considerations (e.g., deployment of technologies that could make decisions that unfairly impact specific groups), privacy considerations, and security considerations.
        \item The conference expects that many papers will be foundational research and not tied to particular applications, let alone deployments. However, if there is a direct path to any negative applications, the authors should point it out. For example, it is legitimate to point out that an improvement in the quality of generative models could be used to generate deepfakes for disinformation. On the other hand, it is not needed to point out that a generic algorithm for optimizing neural networks could enable people to train models that generate Deepfakes faster.
        \item The authors should consider possible harms that could arise when the technology is being used as intended and functioning correctly, harms that could arise when the technology is being used as intended but gives incorrect results, and harms following from (intentional or unintentional) misuse of the technology.
        \item If there are negative societal impacts, the authors could also discuss possible mitigation strategies (e.g., gated release of models, providing defenses in addition to attacks, mechanisms for monitoring misuse, mechanisms to monitor how a system learns from feedback over time, improving the efficiency and accessibility of ML).
    \end{itemize}
    
\item {\bf Safeguards}
    \item[] Question: Does the paper describe safeguards that have been put in place for responsible release of data or models that have a high risk for misuse (e.g., pretrained language models, image generators, or scraped datasets)?
    \item[] Answer: \answerNA{} 
    \item[] Justification: the paper poses no such risks.
    \item[] Guidelines:
    \begin{itemize}
        \item The answer NA means that the paper poses no such risks.
        \item Released models that have a high risk for misuse or dual-use should be released with necessary safeguards to allow for controlled use of the model, for example by requiring that users adhere to usage guidelines or restrictions to access the model or implementing safety filters. 
        \item Datasets that have been scraped from the Internet could pose safety risks. The authors should describe how they avoided releasing unsafe images.
        \item We recognize that providing effective safeguards is challenging, and many papers do not require this, but we encourage authors to take this into account and make a best faith effort.
    \end{itemize}

\item {\bf Licenses for existing assets}
    \item[] Question: Are the creators or original owners of assets (e.g., code, data, models), used in the paper, properly credited and are the license and terms of use explicitly mentioned and properly respected?
    \item[] Answer: \answerYes{} 
    \item[] Justification: We properly credited the creators or original owners of assets (e.g., code, data, models), used in the paper and conformed the license and terms.
    \item[] Guidelines:
    \begin{itemize}
        \item The answer NA means that the paper does not use existing assets.
        \item The authors should cite the original paper that produced the code package or dataset.
        \item The authors should state which version of the asset is used and, if possible, include a URL.
        \item The name of the license (e.g., CC-BY 4.0) should be included for each asset.
        \item For scraped data from a particular source (e.g., website), the copyright and terms of service of that source should be provided.
        \item If assets are released, the license, copyright information, and terms of use in the package should be provided. For popular datasets, \url{paperswithcode.com/datasets} has curated licenses for some datasets. Their licensing guide can help determine the license of a dataset.
        \item For existing datasets that are re-packaged, both the original license and the license of the derived asset (if it has changed) should be provided.
        \item If this information is not available online, the authors are encouraged to reach out to the asset's creators.
    \end{itemize}

\item {\bf New assets}
    \item[] Question: Are new assets introduced in the paper well documented and is the documentation provided alongside the assets?
    \item[] Answer: \answerYes{} 
    \item[] Justification: We communicated the details of the dataset/code/model as part of their submission.
    \item[] Guidelines:
    \begin{itemize}
        \item The answer NA means that the paper does not release new assets.
        \item Researchers should communicate the details of the dataset/code/model as part of their submissions via structured templates. This includes details about training, license, limitations, etc. 
        \item The paper should discuss whether and how consent was obtained from people whose asset is used.
        \item At submission time, remember to anonymize your assets (if applicable). You can either create an anonymized URL or include an anonymized zip file.
    \end{itemize}

\item {\bf Crowdsourcing and research with human subjects}
    \item[] Question: For crowdsourcing experiments and research with human subjects, does the paper include the full text of instructions given to participants and screenshots, if applicable, as well as details about compensation (if any)? 
    \item[] Answer: \answerNA{} 
    \item[] Justification: does not involve crowdsourcing nor research with human subjects
    \item[] Guidelines:
    \begin{itemize}
        \item The answer NA means that the paper does not involve crowdsourcing nor research with human subjects.
        \item Including this information in the supplemental material is fine, but if the main contribution of the paper involves human subjects, then as much detail as possible should be included in the main paper. 
        \item According to the NeurIPS Code of Ethics, workers involved in data collection, curation, or other labor should be paid at least the minimum wage in the country of the data collector. 
    \end{itemize}

\item {\bf Institutional review board (IRB) approvals or equivalent for research with human subjects}
    \item[] Question: Does the paper describe potential risks incurred by study participants, whether such risks were disclosed to the subjects, and whether Institutional Review Board (IRB) approvals (or an equivalent approval/review based on the requirements of your country or institution) were obtained?
    \item[] Answer: \answerNA{} 
    \item[] Justification:  Our paper does not involve study participants.
    \item[] Guidelines:
    \begin{itemize}
        \item The answer NA means that the paper does not involve crowdsourcing nor research with human subjects.
        \item Depending on the country in which research is conducted, IRB approval (or equivalent) may be required for any human subjects research. If you obtained IRB approval, you should clearly state this in the paper. 
        \item We recognize that the procedures for this may vary significantly between institutions and locations, and we expect authors to adhere to the NeurIPS Code of Ethics and the guidelines for their institution. 
        \item For initial submissions, do not include any information that would break anonymity (if applicable), such as the institution conducting the review.
    \end{itemize}

\item {\bf Declaration of LLM usage}
    \item[] Question: Does the paper describe the usage of LLMs if it is an important, original, or non-standard component of the core methods in this research? Note that if the LLM is used only for writing, editing, or formatting purposes and does not impact the core methodology, scientific rigorousness, or originality of the research, declaration is not required.
    \item[] Answer: \answerNo{} 
    \item[] Justification: core method development in this research does not involve LLMs as any important, original, or non-standard components.
    \item[] Guidelines:
    \begin{itemize}
        \item The answer NA means that the core method development in this research does not involve LLMs as any important, original, or non-standard components.
        \item Please refer to our LLM policy (\url{https://neurips.cc/Conferences/2025/LLM}) for what should or should not be described.
    \end{itemize}

\end{enumerate}

\newpage
\appendix
\section{Dense Semantic Retrieval and Ranking}\label{app:dense_semantic}
In the previous section, we derived the spatial candidate set $C_s$ and the spatial relevance score $f_s$. Now, we focus on obtaining the semantic candidate set $C_k$ and the semantic relevance score $f_k$ .

Given a query $q$ , we define the semantic candidate set $C_k(q)$ as:
\begin{equation}
C_k(q) = \{y \mid c_k(y, q) \leq 0, \forall c_k \in \mathcal{C}_k(q)\} ,
\end{equation}
where:
\begin{itemize}[leftmargin=*]
\item $c_k(y, q)$ is a constraint function that filters out spatial objects not satisfying the semantic intent of the query.
\item $\mathcal{C}_k(q)$ is the set of all semantic constraints (e.g., topic matching, category relevance).
\end{itemize}

Each spatial object is associated with textual descriptions, including names, reviews, and additional metadata. However, these descriptions often contain irrelevant or verbose details that may obscure meaningful information. To address this, we use an LLM-based masking function $\mathcal{M}_k$ to remove spatially redundant information and retain only semantically relevant content. The resulting texts are then encoded into a dense embedding space by a text encoder $\mathcal{E}$. Specifically, given a spatial object text description $d_t$,user query $q$, the filtered text representation is:
\begin{equation}
v_{t,k} = \mathcal{E}(\mathcal{M}_k(d_t))\quad v_{q,k} = \mathcal{E}(\mathcal{M}_k(q)).
\end{equation}

The semantic relevance score is then computed using cosine similarity:
\begin{equation}
f_k = \frac{v_{q,k} \cdot v_{t,k}}{ \parallel v_{q,k}\parallel\parallel v_{t,k}\parallel }.
\end{equation}
This score quantifies how well the spatial object aligns with the query's semantic intent, irrespective of spatial factors.

\section{Implementation Details}
\subsection{Semantic Parsing for Spatial Database Query}
For the geometry objects referenced in user queries, Spatial-RAG initially interacts with the spatial database to locate and match the described objects, such as specific points (e.g., a restaurant), roads, or defined areas and subsequently retrieves the pertinent geometrical data. In scenarios where the specified geometrical object does not exist pre-mapped in the database, Spatial-RAG is designed to construct a temporary geometric object. This temporary object serves as a stand-in to facilitate spatial queries based on the user's descriptive input.
This approach allows Spatial-RAG to handle dynamic spatial inquiries efficiently, even when direct matches are not immediately found within the existing database entries. By creating temporary geometrical representations, Spatial-RAG ensures that all spatial queries are processed accurately, maintaining the integrity and effectiveness of the system in delivering precise spatial information and responses.


Functionally, the same outcome might be achieved through different means, for example, searching for a restaurant near a street could involve searching within a buffered polyline or creating a polygon enclosing the polyline and searching within it. Such flexibility in the system implies various methods to achieve the same goal.
This flexibility, however, poses a challenge if the LLM is tasked with generating a complete query directly, as it might lead to the production of hallucinatory, incorrect, or inexecutable code due to confusion or excessive complexity in interpreting spatial data. By structuring the process such that the LLM first identifies the geometry, then determines the function in a step-by-step manner, we mitigate the risks associated with generating errant queries. 
\subsection{Dense Retrieval}
While spatial databases address spatial constraints based on the query and spatial database, the actual scenario may be complex, for instance, a hotel may be far from the airport on the map but provide a shuttle, which makes it spatially more convenient than a hotel closer but do not provide a shuttle. 
Each spatial object is accompanied by textual descriptions, such as names and reviews. However, the text often contains verbose and irrelevant details that hinder effective decision-making. 
Moreover, for areas with a high density of POIs that meet spatial requirements, it becomes impractical to input all the text information into an LLM (Large Language Model). To manage the data volume and improve relevance to specific queries, these descriptions are summarized across two perspectives:  spatial requirements and semantic requirements. 
We utilize an LLM to preprocess and summarize spatial objects' reviews offline, storing the results in the database for future comparison.
Similarly, the user queries are dynamically extracted during the online processing stage.

\section{Prompt list}
\lstset{
  basicstyle=\ttfamily\small,   
  numbers=left,                 
  numberstyle=\tiny\color{gray},
  frame=single,                 
  breaklines=true,              
  columns=fullflexible,         
  keepspaces=true,              
  xleftmargin=2em,              
  framexleftmargin=2em,         
}

\subsection{Prompt: Spatial Information Extraction}
Extract spatial information from user queries, including object geometry type (Point, Polyline (Route), or Polygon (Region)), region name, distance, and buffer distance.
\begin{lstlisting}
Analyze the following user query and extract spatial information: "{user_query}"

Current location context:
- Number of location points: {location_count}
- Multiple points detected: {is_multi_point}

First, determine the spatial query type based on these rules:
1. For single location point ({location_count == 1}):
   - Use Region-based if query explicitly mentions a region
   - Otherwise, use Point-based
   
2. For exactly two points ({location_count == 2}):
   - Use Route-based if query suggests path/route between points
   - Otherwise, fall back to Point/Region based rules
   
3. For multiple points ({location_count > 2}):
   - Only use Point-based or Region-based

Query types:
1. Point-based:
   - For "nearby" or "close": 1km in dense areas
   - For "walking distance": 2km
   - For "not too far": 3km
   
2. Route-based:
   - ONLY available with exactly 2 points
   - For walking routes: 1000m buffer
   - For general routes: 2000m buffer
   - For scenic/exploration: 3000m buffer
   - Consider terms: "route", "path", "between", "from...to", "along"
   
3. Region-based:
   - ONLY if query explicitly mentions these regions:
   Community/Sub-region names: {', '.join(region_names['nta_names'])}
   Borough names: {', '.join(region_names['boro_names'])}
   - Do NOT infer regions from landmarks

Return in strict JSON format:
{
    "query_type": "point" | "route" | "region",
    "region": "matched region name or null",
    "distance_km": number or null,
    "buffer_distance": number or null,
}
\end{lstlisting}

\subsection{Prompt: Semantic Intent Extraction}
Extract semantic intent from user queries, including spatial and nonspatial preference.
\begin{lstlisting}
Analyze the following user query and extract constraints: "{user_query}"

First, determine the main purpose of the query by identifying key terms and context:

Restaurant (R) keywords and contexts:
- Direct terms: "restaurant", "food", "eat", "dining", "meal", "cuisine"
- Food types: "Chinese", "Thai", "Mexican", "Italian", "sushi", etc.
- Meal times: "breakfast", "lunch", "dinner", "brunch"
- Dining related: "menu", "dishes", "chef", "reservation"
- Even if staying at a hotel, if asking about food/dining, it's Restaurant (R)

Hotel (H) keywords and contexts:
- Must be explicitly looking for accommodation
- Direct terms: "hotel", "stay", "accommodation", "room", "book"
- Price per night (e.g., "$200/night")
- Hotel names (e.g., "Hyatt", "Marriott")
- Mentioning a hotel as location reference is NOT H type

Attraction (A) keywords and contexts:
- Direct terms: "visit", "see", "tour", "explore"
- Places: "museum", "park", "gallery", "theater"
- Activities: "sightseeing", "show", "performance"

Important rules:
1. Focus on what the user is ASKING FOR, not what they mention
2. If user mentions staying at a hotel but asks about restaurants, type is R
3. If query is about food/dining/restaurants, type must be R
4. Location references (e.g., "near Hotel X") don't determine type

For each constraint type, extract complete sentences that describe the requirements:


1. Spatial constraints: Where they want to go
   Example: "near Times Square" or "in the Upper West Side area"

2. User constraints: What specific requirements or preferences they have
   Example: "family-friendly restaurant with reasonable prices around $30 per person"

Please return strict JSON format without any comments:
{
    "type": "R/H/A",
    "spatial_constraints": "complete sentence describing location requirements or null",
    "user_constraints": "complete sentence describing user preferences and requirements or null"
}
\end{lstlisting}

\subsection{Prompt: Result Reranking}
Rerank retrieved location results based on user query constraints.
\begin{lstlisting}
As a local recommendation expert, please rank the following places based on user query constraints.

User Query Constraints:
- Spatial Constraints: {query_constraints['spatial_constraints']}
- User Preferences: {query_constraints['user_constraints']}

Candidate Places:
{json.dumps(places, ensure_ascii=False, indent=2)}

Please analyze how well each place matches the user constraints and return a sorted list of places.
Return format should be a JSON array containing sorted indices.
Only return the index array, e.g., [2,0,1,3] means the 3rd place is the best match, followed by 1st, 2nd, and 4th places.
Note: Must return indices for all places, array length should equal input place count ({len(places)}).
\end{lstlisting}

\section{Limitations}
We employ the ChatGPT API for direct LLM-based detection of target and reference objects. While this approach provides immediate functionality, future work will investigate fine-tuning a specialized language model to enhance performance for this specific task.

\section{Additional case studies}
\subsection{Case 1}
\textbf{User Query:} \\
\textit{Going to be in Manhattan for a Broadway show this weekend and am looking for a recommendation for a good restaurant in the theatre district that is reasonably priced (\$30 to \$50 per person), good food and not a tourist trap, would love to avoid the hotel restaurants and experience a real NY experience. Does such a place exist? Someone suggested Carmine's but they are totally booked. (like all kinds of food) Any suggestions would be greatly appreciated - picking one out of the 5000+ listed is a daunting task.}

\textbf{Model Processing Summary:} \\
The system first determines that the \textbf{reference object type} is \textbf{polygon}, and correctly identifies the \textbf{region as Manhattan}. It detects that the \textbf{target object type} is a restaurant \textbf{ point}. Based on these constraints, a SQL query is constructed to retrieve \textbf{532 candidate locations} from the database.

The \textbf{spatial requirement} is: \\
\textit{"am looking for a recommendation for a good restaurant in the theatre district"}

The \textbf{semantic requirements} are: \\
\textit{"reasonably priced (\$30 to \$50 per person), good food and not a tourist trap, avoid hotel restaurants, real NY experience"}

The system computes vector similarities between the query and the reviews of all 532 candidates, considering both spatial and semantic relevance. The top 20 results based on combined score are passed to an {LLM reranker}, and the final \textbf{top-1 recommendation} is:

\textbf{West Bank Cafe, 407 W 42nd St, New York City, NY 10036-6808}

\textbf{Reasoning:}
\begin{itemize}
    \item \textbf{Spatial match:} West Bank Cafe is located in the Theatre District of Manhattan, satisfying the user’s regional constraint. Its location aligns well with proximity to Broadway shows.
    \item \textbf{Semantic alignment:} The restaurant is described as having reasonable pricing and good food, with a local, non-touristy atmosphere. These attributes align with the user's preferences for affordability and an authentic New York experience.
\end{itemize}

\subsection{Case 2}

\textbf{User Query:} \\
\textit{Going to NY in May, it is my gf's birthday on our last night – and I would be looking for a nice restaurant to spoil her :)! We are staying in Times Square – so would like somewhere close by! I would like either an American or Italian cuisine! Any advice would be much appreciated.}

\textbf{Model Processing Summary:} \\
The model identifies the query as a \textbf{point} spatial search with a radius of \textbf{1.0 km} around \textbf{Times Square}. The \textbf{target object type} is recognized as a restaurant \textbf{point}. Based on these criteria, a SQL query returns \textbf{119 candidate locations} from the database.

The \textbf{spatial requirement} is: \\
\textit{"We are staying in Times Square – so would like somewhere close by!"}

The \textbf{semantic requirements} are: \\
\textit{"it is my gf's birthday on our last night"}, \\
\textit{"I would like either an American or Italian cuisine!"}

The system computes vector similarities based on spatial and semantic constraints. The top 20 candidates are re-ranked by an {LLM reranker}, and the final \textbf{top-1 recommendation} is:

\textbf{Pasta Lovers, 142 W 49th St, New York City, NY 10019-6802}

\textbf{Reasoning:}
\begin{itemize}
    \item \textbf{Spatial match:} Pasta Lovers is located within 1 km of Times Square, satisfying the proximity requirement effectively.
    \item \textbf{Semantic alignment:} It offers Italian cuisine, aligning with the user's preference for either American or Italian food. The atmosphere and reviews suggest a cozy dining experience appropriate for a date, though it is not explicitly described as suitable for a birthday celebration.
\end{itemize}

\begin{figure*}[htbp]
  \centering
  \includegraphics[width=\textwidth]{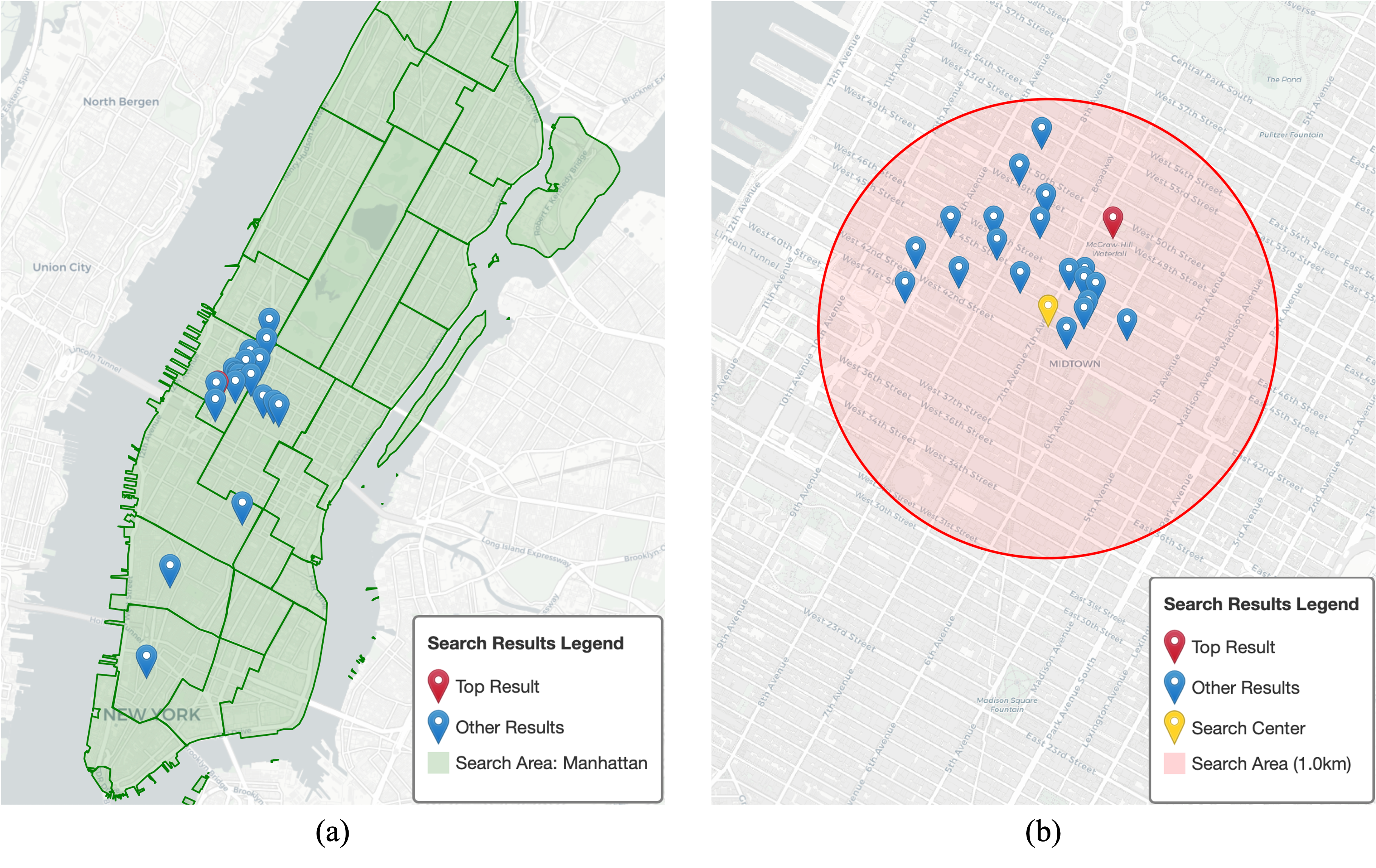}
  \caption{{(a) Case 1; (b) Case 2}}
  \label{fig:case12}
\end{figure*}

\subsection{Case 3}

\textbf{User Query:} \\
\textit{Visiting NYC for the Marathon this weekend. Staying in apartment in Hell's Kitchen area (around 51st and 12th). Anyone have good recommendations for a great breakfast restaurant within an 8–10 block radius? Not worried about eating healthy the day before—looking for filling, traditional American-style breakfast! Also looking for ideas for a good filling dinner for Sunday evening after the race. We're not looking for fine dining, just something good in the area for myself, wife and 16 yo daughter. Thanks in advance for any good recommendations.}

\textbf{Model Processing Summary:} \\
The model identifies the query as a \textbf{point} spatial query with a radius of \textbf{1.5 km}, derived from the “8–10 block” constraint. The \textbf{target object type} is restaurant \textbf{point}, and the approximate location is Hell’s Kitchen near 51st St and 12th Ave. A SQL query retrieves \textbf{140 candidates} from the database based on this spatial condition.

The \textbf{spatial requirement} is: \\
\textit{"Staying in apartment in Hell's Kitchen area (around 51st and 12th), breakfast restaurant within an 8–10 block radius"}

The \textbf{semantic requirements} include: \\
\textit{"traditional American-style breakfast"}, \textit{"filling dinner after the race (Sunday evening), just something good in the area for myself, wife and 16 yo daughter."}

The system calculates vector similarity scores based on spatial and semantic requirements, and selects the top 20 results. These are re-ranked by an {LLM reranker}, and the final \textbf{top-1 recommendation} is:

\textbf{Galaxy Diner, 665 9th Ave, New York City, NY 10036-3623}

\textbf{Reasoning:}
\begin{itemize}
    \item \textbf{Spatial match:} Galaxy Diner is located within a reasonable walking distance  of the user’s location, satisfying the “8–10 block radius” constraint in Hell’s Kitchen.
    \item \textbf{Semantic alignment:} The diner offers traditional American breakfasts like pancakes, eggs, and bacon, aligning with the user’s request for filling, non-healthy food before the race. It also accommodates a casual family atmosphere.
\end{itemize}

\subsection{Case 4}
\textbf{User Query:} \\
\textit{We are staying in Midtown, so figured it would be easier to go somewhere close by, but I'm not opposed to somewhere a little further away. Looking for a restaurant where we can go to dinner and dress up but not have it be a crazy price. I understand that a "dress up" place isn't usually cheap, but something on the lower end of the "dressy" price scale would be great :) Thanks!}

\textbf{Model Processing Summary:} \\
The model correctly identifies the  \textbf{reference object type} as \textbf{polygon} (\textbf{Midtown}). A SQL query retrieves \textbf{97 candidates} from the database based on this spatial condition.

The \textbf{spatial requirement} is: \\
\textit{"Staying in Midtown, somewhere close by, but not opposed to somewhere a little further away."}

The \textbf{semantic requirement} is: \\
\textit{"restaurant where we can go to dinner and dress up but not have it be a crazy price, lower end of the 'dressy' price scale."}

The model calculates spatial and user constraint similarities, ranks the top 20 results, and applies an {LLM reranker}. The final \textbf{top-1 recommendation} is:

\textbf{Nocello, 257 W 55th St, New York City, NY 10019-5232}

\textbf{Reasoning:}
\begin{itemize}
    \item \textbf{Spatial match:} Nocello is within a reasonable walking distance of Midtown, fulfilling the requirement of being nearby but not necessarily adjacent. It’s described as a short walk from Restaurant Row and near Broadway, aligning well with the user’s open spatial boundary.
    \item \textbf{Semantic alignment:} The restaurant is described as offering excellent food at a reasonable price, making it suitable for a "dress-up" dinner without the high-end cost. This aligns with the user's goal of finding something elegant but affordable.

\end{itemize}

\begin{figure*}[htbp]
  \centering
  \includegraphics[width=\textwidth]{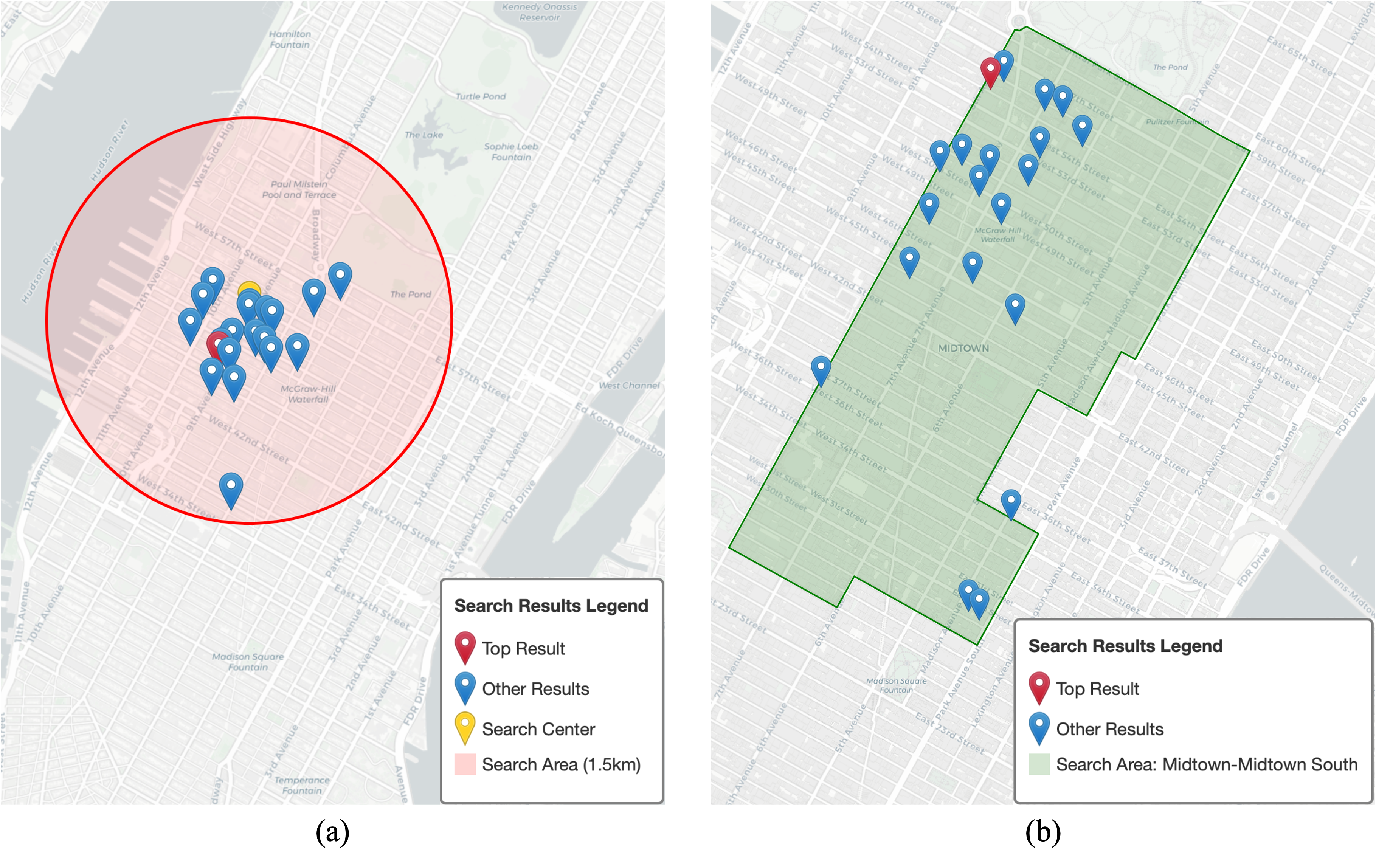}
  \caption{{(a) Case 3; (b) Case 4}}
  \label{fig:case34}
\end{figure*}

\subsection{Case 5}

\textbf{User Query:} \\
\textit{Where are some good places to eat breakfast? We are staying in the southern tip of Manhattan near Battery Park, but it doesn't have to be contained to just that area. We are very open to ideas, but are going to be avoiding McDonalds, BK, etc. Could be larger restaurants but also would like to visit a few small places and would like to be able to sit and eat outside.}

\textbf{Model Processing Summary:} \\
The model correctly identifies the  \textbf{reference object type} as \textbf{polygon} (\textbf{battery park city-lower manhattan}).  A SQL query retrieves \textbf{14 candidates} from the database based on this spatial condition.

The \textbf{spatial requirement} is: \\
\textit{"staying in the southern tip of Manhattan near Battery Park, but it doesn't have to be contained to just that area."}

The \textbf{semantic requirements} are: \\
\textit{"avoiding McDonalds, BK, etc. Could be larger restaurants but also would like to visit a few small places and would like to be able to sit and eat outside."}

The model ranks the top 10 by spatial and user relevance, and performs {LLM reranking}. The final \textbf{top-1 recommendation} is:

\textbf{Stone Street Tavern, 52 Stone St, New York City, NY 10004-2604}

\textbf{Reasoning:}
\begin{itemize}
    \item \textbf{Spatial match:} The restaurant is located a short walk from Battery Park in the financial district, consistent with the user’s desire to explore areas nearby but not strictly limited to Battery Park. The cobblestone street setting and access to outdoor space align well with the user's spatial intent.
    \item \textbf{Semantic alignment:} Stone Street Tavern offers outdoor seating and avoids fast-food chains. It provides a casual and local dining experience with bench-style outdoor tables, fitting the user’s interest in both large and small sit-down places for breakfast.
\end{itemize}

\subsection{Case 6}

\textbf{User Query:} \\
\textit{We have tickets for the Saturday performance of War Horse at the Vivian Beaumont Theater at Lincoln Center. I would appreciate a recommendation for a reasonable pre-theater dinner.}

\textbf{Model Processing Summary:} \\
The model correctly identifies the \textbf{reference object type} as \textbf{point} query, as the user provides a specific point of interest (\textbf{Vivian Beaumont Theater}) without mentioning a formal region. A walking distance of 2.0 km is assumed for pre-theater dining.  A SQL query retrieves \textbf{255 candidates} from the database based on this spatial condition.

The \textbf{spatial requirement} is: \\
\textit{"Vivian Beaumont Theater at Lincoln Center" – implying a location within walking distance}

The \textbf{semantic requirements} are: \\
\textit{"reasonable pre-theater dinner"} — suggesting affordability and suitable timing for a theater schedule.

The model reranks the top 20 via {LLM reranking}. The final \textbf{top-1 recommendation} is:

\textbf{Bar Boulud, 1900 Broadway, New York City, NY 10023-7004}

\textbf{Reasoning:}
\begin{itemize}
    \item \textbf{Spatial match:} Bar Boulud is located directly across from Lincoln Center, fulfilling the proximity constraint perfectly for a pre-theater dinner.
    \item \textbf{Semantic alignment:} The restaurant is known for accommodating theatergoers and offers timing suitable for pre-show dining. However, user reviews mention that while the food and service are excellent, prices may be higher than what the user considers “reasonable”.
\end{itemize}

\begin{figure*}[htbp]
  \centering
  \includegraphics[width=\textwidth]{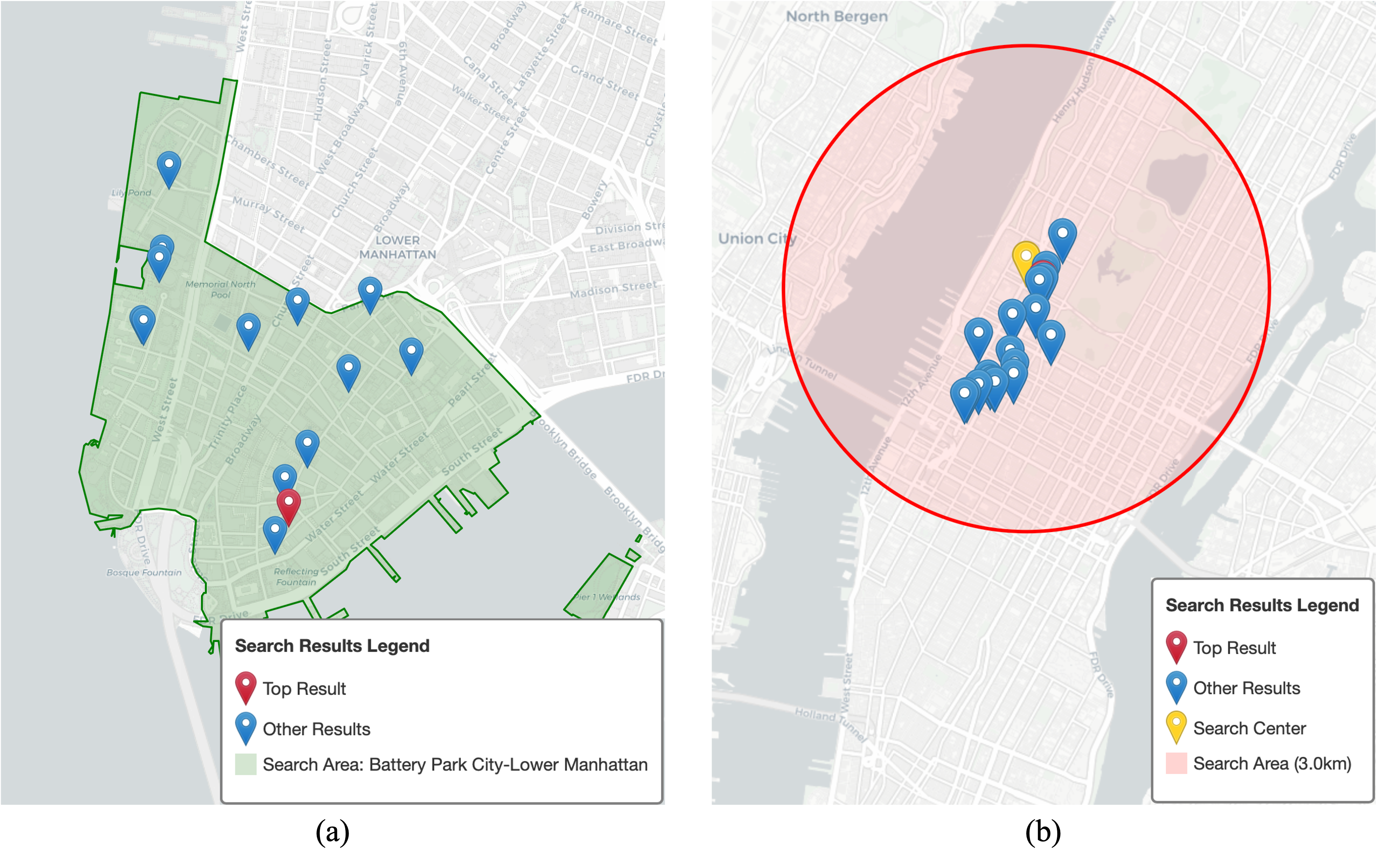}
  \caption{{(a) Case 5; (b) Case 6}}
  \label{fig:case56}
\end{figure*}

\subsection{Case 7}

\textbf{User Query:} \\
\textit{I want to walk the Highline on my forthcoming trip to NY, and wanted some recommendations for a good spot for a sit down lunch somewhere close to one of the exits off the Highline. Will be starting north to south. Don’t mind what type of food, just nice atmosphere required. Have looked at MenuPages but so many restaurants not sure if anyone has any particular favourites?}

\textbf{Model Processing Summary:} \\
The model correctly identifies the \textbf{reference object type} as \textbf{polyline}. This is inferred from the user's mention of walking from the north to south along the Highline, with interest in locations near the path (Highline exits). Two coordinates representing the start and end of the route are provided. The buffer distance is set to 500 meters.

The \textbf{spatial requirement} is: \\
\textit{"somewhere close to one of the exits off the Highline."}

The \textbf{semantic requirement} is: \\
\textit{"sit down lunch, nice atmosphere required."}

The system extracts 52 candidate restaurants along the route, ranks them by spatial and user constraints, and performs {LLM reranking} on the top 20. The final \textbf{top-1 recommendation} is:

\textbf{Barbuto, 775 Washington St, New York City, NY 10014-1748}

\textbf{Reasoning:}
\begin{itemize}
    \item \textbf{Spatial match:} Barbuto is located directly off the Highline and near one of its southern exits, which aligns precisely with the user’s request for proximity to the walking path.
    \item \textbf{Semantic alignment:} The restaurant offers a sit-down experience with a "fancy but cozy" atmosphere, partially fulfilling the user's request for a "nice atmosphere." However, some reviews describe the food as average, which may not fully satisfy quality expectations.
\end{itemize}

\subsection{Case 8}

\textbf{User Query:} \\
\textit{We arrive Sat night at 10:30 (unfortunately AA changed our flight from an earlier one). We are spending 1 night at RCA at 142 W 49th between 6th and 7th. Sunday we move to Hilton New York at 1335 Ave Of Americas between 53 and 54. We have tickets for a Yankee game at 1:05. (We will also be going to Dizzy's Coca Cola at 9PM.) Could you guys give me some tips on the best way to handle the hectic morning? Leave our bags at RCA? Move them to Hilton for storage so we can get subway to Yankee Stadium? Can you give me some recommendations for dinner that evening - something simple. Thanks!}

\textbf{Model Processing Summary:} \\
The model correctly identifies the \textbf{reference object type} as \textbf{polyline}, given that the user describes movement between two locations: Hilton New York and Yankee Stadium. The user's intention to find a dinner spot along that route aligns with this classification. The buffer distance is set to \textbf{500 meters}.

"The \textbf{spatial requirement} is: \\
\textit{Implied need for proximity to Hilton New York or Yankee Stadium for convenience."}

The \textbf{semantic requirements} are: \\
\textit{"We will also be going to Dizzy's Coca Cola at 9PM." and "recommendations for dinner that evening - something simple."}

The system identifies 50 restaurant candidates within the route buffer, computes user and temporal similarity, and applies {LLM reranking} on the top 20. The final \textbf{top-1 recommendation} is:

\textbf{Good Enough to Eat, 520 Columbus Ave, Frnt A, New York City, NY 10024-3404}

\textbf{Reasoning:}
\begin{itemize}
    \item \textbf{Spatial match:} This place is within the buffer range, and the restaurant is reasonably accessible and could fit into the evening schedule before Dizzy’s Coca Cola.
    \item \textbf{Semantic alignment:} The user requested a simple dinner. The reviews highlight casual meals like pancakes, BLT omelets, and quick service, matching the preference for something simple—though most mentions are for breakfast/brunch.
\end{itemize}

\begin{figure*}[htbp]
  \centering
  \includegraphics[width=\textwidth]{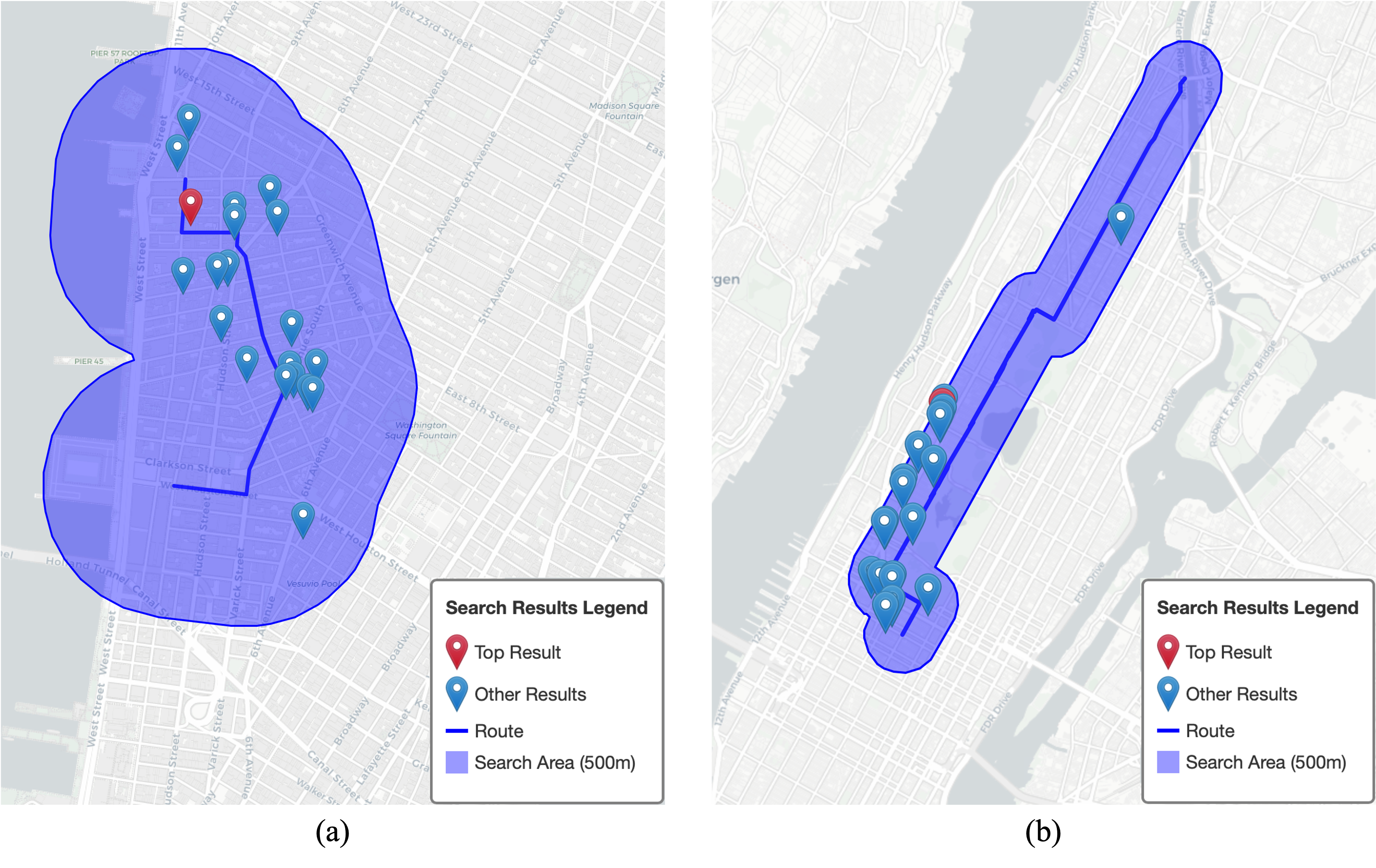}
  \caption{{(a) Case 7; (b) Case 8}}
  \label{fig:case78}
\end{figure*}

\end{document}